\documentclass[<options>]{elsarticle}
\biboptions{sort&compress}

\usepackage{graphicx}
\usepackage{caption}
\usepackage{threeparttable}
\usepackage{subfigure}
\usepackage{tabularx}
\usepackage{booktabs}
\usepackage{lipsum}
\usepackage{multirow}
\usepackage[usenames,dvipsnames]{color}
\usepackage{colortbl}
\definecolor{mygray}{gray}{.9}

\usepackage{amsmath,bm}
\usepackage[figuresright]{rotating}

\usepackage{algorithm,algpseudocode}
\usepackage{amsmath}
\usepackage{amsfonts}
\usepackage{amssymb}

\usepackage[normalem]{ulem}

\journal{Applied Energy}
\begin{document}
\begin{frontmatter}




\title{A Novel Koopman-Inspired Method for the Secondary Control of Microgrids with Grid-Forming and Grid-Following Sources}


\author[]{Xun Gong}
\author[]{Xiaozhe Wang\corref{cor1}}

\cortext[cor1]{Corresponding author}
\fntext[cor1]{This work was supported by the Fonds de Recherche du Quebec-Nature et technologies under Grant FRQ-NT PR-298827 and NSERC ALLRP 571554 - 21.}

\address{Department of Electrical and Computer Engineering, McGill University, 3480 Rue University, Montreal,H3A 0E9, Quebec, Canada}
            
\begin{abstract}
This paper proposes an online data-driven Koopman-inspired identification and control method for microgrid secondary voltage and frequency control. Unlike typical data-driven methods, the proposed method requires no warm-up training yet with guaranteed bounded-input-bounded-output (BIBO) stability and even \color{black} asymptotic \color{black} stability under some mild conditions. 
The proposed method estimates the Koopman state space model adaptively 
so as to perform effective secondary voltage and frequency control that can handle microgrid nonlinearity and uncertainty. 
Case studies in the 4-bus and 13-bus microgrid test systems 
(with grid-forming and grid-following sources) demonstrate the effectiveness and robustness of the proposed identification and control method subject to the change of operating conditions and large disturbances (e.g.,  microgrid mode transition\color{black}s\color{black},  generation/load variations) even with measurement noises and time delays. 
\end{abstract}







\end{frontmatter}

\section{Introduction}
The microgrids (MGs) are small local grids that can disconnect from the bulk grid to operate independently. The MGs facilitate the integration of sustainable distributed energy resources (DERs) like wind, solar as well as energy storage. Nonetheless,
the DERs are interfaced with microgrids by power converters, making MGs low-inertia or even inertia-less \cite{Schneider2019,TFMG2020}. In addition, MGs are characterized by frequency-voltage dependence due to low X/R ratios as opposed to conventional power systems \cite{TFMG2020,Zhang2016}. Therefore, the frequency and voltage of MGs tend to experience coupled large deviations subject to volatile operation conditions of generation and load, transition\color{black}s \color{black} between islanded and grid-connected modes, etc. 

The hierarchical control is commonly adopted to maintain the MG's voltage and frequency stability.   
The hierarchical control includes primary control at the individual DER level, and secondary/tertiary control at the systemwide level. 
Even though the droop-based primary control 
at individual DERs can 
coordinate the power of DERs in a decentralized manner and improve the local stability, the frequency and voltage deviation at the system level may not be eliminated by merely using primary control \cite{Simpson2015}. The system stability may even be compromised when the droop gains are improperly designed to high values \cite{Majumder2010}. Hence, the secondary control is essential to achieve stable voltage and frequency restoration. \color{black}  

The scope of the paper lies in the secondary control, aiming to restore frequency and voltage for islanded MGs under larger disturbances  or 
MGs 
in mode transition\color{black}s\color{black} (e.g., from grid-connected to islanded). 
The secondary control of MGs can be classified as model-based and model-free. 
There have been many research papers on model-based control. For example, the small-signal models have been used in \cite{Zhang2016, HONARVARNAZARI2012823} to  
regulate droop gains and improve the systemwide small-signal stability. To handle large disturbance, multi-agent distributed cooperative control with feedback linearization was proposed in \cite{Bidram2014,Bidram2013} to deal with the nonlinearity. 
Despite advancement, all the aforementioned methods rely heavily on accurate physical models that may not always be available to MG operators due to \color{black}time-varying topologies and operating conditions, as well as \color{black} high uncertainty introduced by volatile renewables. 

To relax the pre-knowledge of accurate models, researchers have designed various model-free control methods. 
A common model-free method is Proportional and Integral (PI) control \cite{Simpson2015,Ma2021,Ahumada2016}, which nevertheless may  
lack online adaptiveness to compensate for uncertainty. Besides, 
the MG may suffer from high starting overshoot, high sensitivity to controller gains, and sluggish response to disturbances if the PI control is not properly tuned \cite{Ma2021}. Another category of MG secondary control method is the averaging/consensus-based secondary droop control \cite{Nutkani2015,Lu2015, LuLY2018} that targets on accurate power sharing in quasi steady-state rather than voltage and frequency stability under large disturbances. To improve systemwide voltage and frequency stability under both small and big disturbances, machine learning based methods were proposed \cite{Ma2021,Jafari2018,Shen2019,Amoateng2017,Lin2021,Shayeghi2019} for secondary voltage and frequency control. \color{black}However, the universal learning machines such as artificial neural network \color{black}(ANN) \color{black} and reinforcement learning (RL) may lack physical interpretability and thus reliability of representing the system’s dynamics in diverse topologies and operating conditions. Obtaining \color{black}adequate \color{black} offline training data that can sufficiently represent the system dynamics is challenging too. 

Moreover, individual DERs can be either controllable (e.g., energy storage systems (ESSs), renewable energy with ESSs), or non-controllable (e.g., renewable generation operating under maximum power point tracking (MPPT)) at the secondary level. They possess diverse modes of primary control (e.g., conventional isochronous grid-forming, power-based grid-forming and grid-following). 
The resulting model complexity may affect the performance of the secondary control. 
Yet all the aforementioned works (both model-based and model-free) on secondary voltage and frequency control assume that all DERs in islanded MGs work under the grid-forming or voltage control mode \cite{Majumder2010,Bidram2014,Bidram2013,Ma2021,Ahumada2016,Shen2019,Amoateng2017,Lin2021,R.Zhang2019}. 
However, in existing MGs, the mix of grid-forming and grid-following control with diverse control structures and parameters introduces uncertainty that challenges MG secondary control. Particularly, when large disturbances occur, the interaction among diverse grid-forming and grid-following converters and the dynamics of the affiliated phasor-locked loops (PLLs) may \color{black} deteriorate the system stability and control performance. 
\color{black} 
\color{black}

In this paper, we propose a new data-driven secondary voltage and frequency control method for MGs with both grid-forming and grid-following DERs. The method is able to handle MG nonlinearity and uncertainty (e.g., MG mode transition\color{black}s \color{black} from grid-connected to islanded, generation and load variations) in an adaptive data-driven fashion. The proposed method requires no offline training and uses only a small window of phasor angle and voltage data from synchrophasors (e.g., micoPMUs) at the DER output ends. In the proposed method, Koopman operator theory \cite{Proctor2018} is leveraged to convert the nonlinear dynamical system into a linear one under 
\color{black} Koopman embedding mapping. As such, 
the system can be identified and controlled with mature and powerful linear system techniques.  
Particularly, we tailor \color{black} the OKID (Observer Kalman-filter IDentification)-based algorithm 
so that the Koopman-based linear dynamical system 
can be identified optimally. Then, the discrete-time linear quadratic regulator (LQR) is applied to the identified Koopman-based linear dynamical system 
with well-characterized stability properties.
It is noteworthy that  M. Korda et al \cite{KORDA2018297}  utilized Koopman operator control for power system transient stability and control, while the method required offline training data. Besides, the identification based on 
the brute-force 
least-squares estimation, 
could lead to unsatisfactory identification results. \color{black} Gong et al \cite{Gong2022} presented a combined application of the Koopman operator and identification method for MG secondary control. However, the method assumes that the droop parameters of DERs are known by the secondary controller, whereby the control matrix in the Koopman state space can be directly obtained. In this paper, we lift the assumption that the local control mechanism and parameters are fully unknown. \color{black}
\color{black}
In short, the advantages of the proposed method are summarized as below:

\noindent (i) The proposed Koopman-inspired enhanced OKID method can \color{black} help \color{black} identify the system dynamics accurately and adaptively \color{black} due to the capacity of dealing with nonlinearity and uncertainty under large disturbances.\color{black}

\noindent(ii) The proposed Koopman-inspired identification and control method 
 is purely data-driven using only a small window of synchrophasor data. \color{black} It requires \color{black} no knowledge of network information and primary controllers\color{black}, and no offline training.


\noindent(iii) The MG system with the proposed Koopman-inspired identification and control is guaranteed to be bounded-input-bounded-output (BIBO) stable. On top of the BIBO stability, the sufficient condition under which the MG system is asymptotically stable is also developed. 

\noindent(iv) The proposed control method is robust to measurement noises and time delays as tested in numerical studies.

The remainder of the paper is organized as follows: Section II describes the MG hierarchical control and the interfaces between secondary and primary control. Section III details the proposed Koopman-inspired identification and control method. Section IV presents case studies for validation. Section V concludes the paper.

\begin{figure}[!tb]
\centering
  \includegraphics[width=0.6\linewidth]{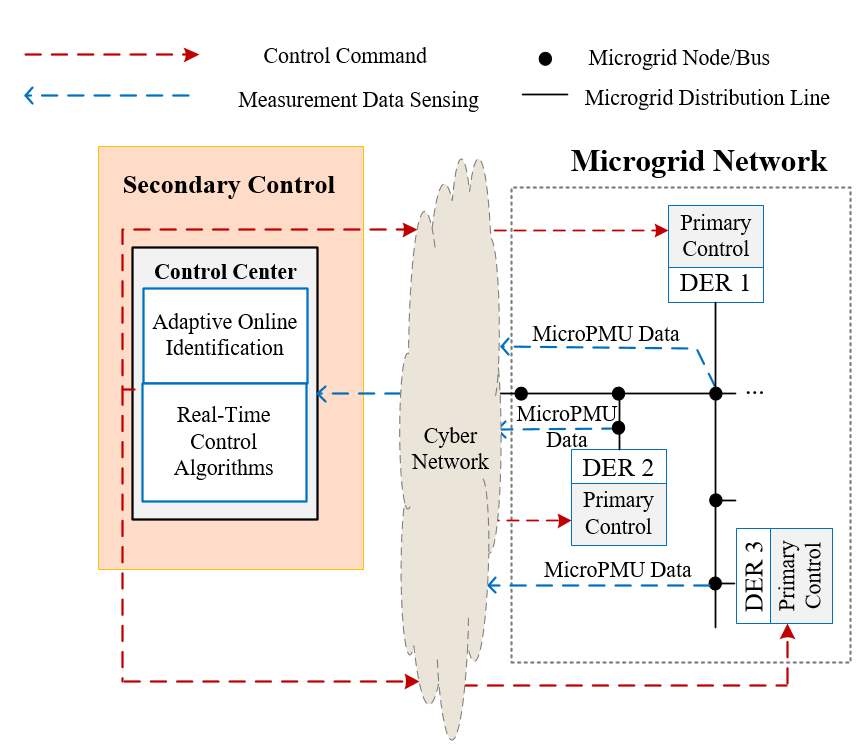}
  \caption{\color{black} Microgrid control architecture}
  \label{MGarch}
\end{figure}

\section{Microgrid System Description}
A MG can be controlled hierarchically with the secondary and primary control as shown in Fig. \ref{MGarch}. The primary controllers enable fast response of the individual DERs to guarantee local stability, while the secondary controller globally dictates the primary controllers of controllable DERs according to the data collected from microPMUs, whereby the systemwide interaction dynamics of MGs can be \color{black}handled \color{black} and the voltage and frequency can be restored.

\begin{figure}[!tb]
\centering
  \includegraphics[width=0.501\linewidth]{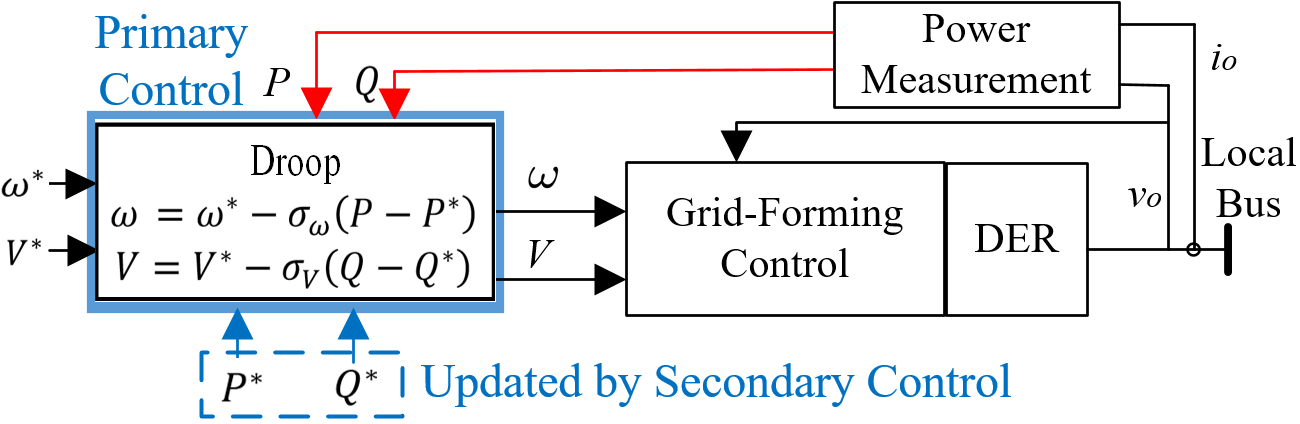}
  \includegraphics[width=0.49\linewidth]{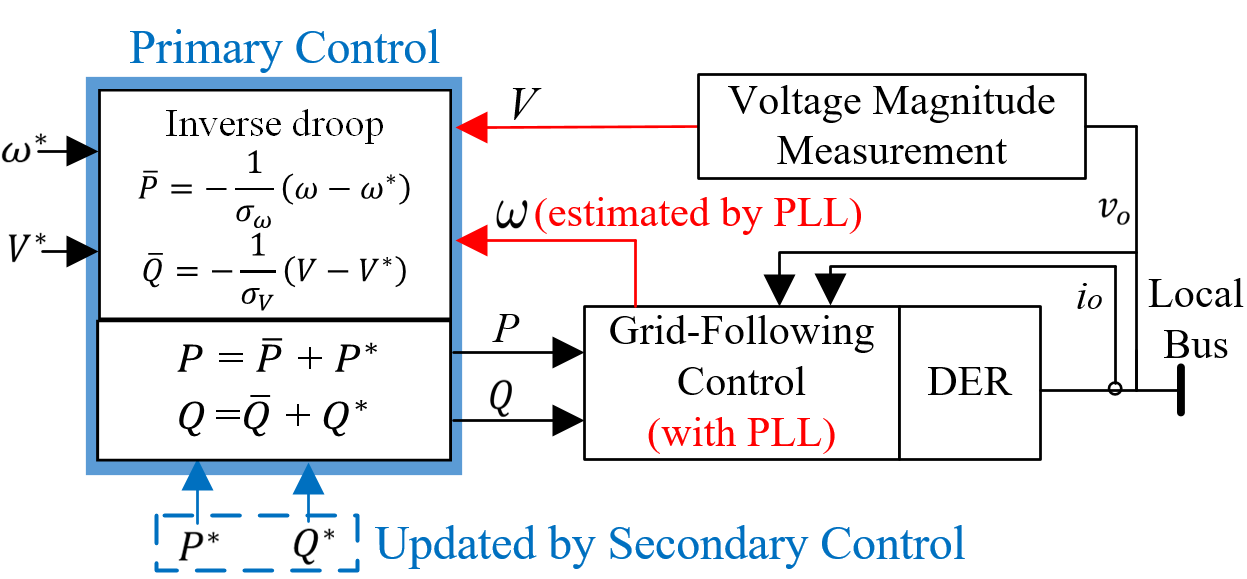}\\
  (a)  \qquad \qquad \qquad \qquad \qquad \qquad \qquad (b) 
  \caption{Different primary control modes: (a) droop-based grid-forming control; (b) inverse-droop-based grid-following  control.}
  \label{fig: primarycontrol}\end{figure}

The local primary control modes can be different, such as grid-forming or grid-following \cite{NREL2020}. In an islanded MG or future power system without synchronous generators, at least one DER is required to work under \color{black} the \color{black} grid-forming mode to actively form the grid voltage and frequency; then the rest of DERs can remain operating under \color{black} the \color{black} grid-following modes \cite{NREL2020}. 
As discussed in \cite{6200347,Kroposki2016}, grid-forming DERs \textit{define} the voltage magnitude and frequency. In contrast, grid-following DERs \textit{follow} the measured frequency and voltage magnitude in the grid via PLL, which represents the prevalent type of control strategy for grid-connected PV and wind converters in existing power grids. \color{black}
As droop is commonly used in MGs, we consider the droop-based grid-forming \cite{GFM_PLL2019,Wei2020} and inverse-droop-based grid-following control \cite{Ryan2021}. 
Specifically, consider a DER at the bus $i$. As shown in Fig. \ref{fig: primarycontrol}(a), the droop for grid-forming converter control is defined as:
\begin{small}
\begin{equation}
\begin{aligned}
Droop:\left[ \begin{array}{l}{\omega_{i}-\omega^*_{i}}\\
{V_{i}-V^*_{i}}
\end{array} \right]&=
\left[ \begin{array}{l}
{-\sigma_\omega(P_{i}-P^*_{i})}\\
-\sigma_V(Q_{i}-Q^*_{i})
\end{array} \right]
\label{eq:DROOPS}
\end{aligned}
\end{equation}
\end{small}\noindent
where $\omega_{i}$ and $V_{i}$ denote the frequency  and voltage magnitude for the grid-forming control. 
$\omega^{*}_{i}$ and $ V^{*}_{i}$ are the rated frequency and voltage. The parameters $\sigma_{\omega}$ and $\sigma_{V}$ are frequency and voltage droop gains, respectively. $ P^{*}_{i}$ and $ Q^{*}_{i}$ are the reference power in the droop, which 
can be the steady-state power without the secondary control or an augmented reference power after the secondary control.  
$P_{i}$ and $Q_{i}$ are the active and reactive power, \color{black} which are measured with a low-pass filter embedded in the power measurement block in Fig. \ref{fig: primarycontrol}. The filter is in the form of \cite{Pogaku2007,Hassan2013}:
\begin{small}
\begin{equation}
  P_{i}=\frac{1}{T_fs + 1} P^{(IN)}_{i}, \quad   Q_{i}=\frac{1}{T_fs + 1} Q^{(IN)}_{i}
    \label{eq:gfsc}
\end{equation}
\end{small}\noindent
where $T_f$ is the time constant of the first-order low-pass filter; $P^{(IN)}_{i}$ and $Q^{(IN)}_{i}$ represent the active and reactive power before filtering. Generally, the filter is required to attenuate high-frequency dynamics (e.g., harmonics) and preserve low-frequency dynamics (e.g., sub-synchronous components which can be further managed by secondary control). \color{black}With the secondary control, the reference power $P^*_{i}$ and $Q^*_{i}$ of the droop in Fig. \ref{fig: primarycontrol} was updated in discrete time as: 
\begin{small}
\begin{equation}
  P^{*(+)}_{i}=P^{*}_{i} + \Delta P^{*}_{i}, \quad Q^{*(+)}_{i}= Q^{*}_{i} + \Delta Q^{*}_{i}
    \label{eq:gfsc}
\end{equation}
\end{small}\noindent
To distinguish $P^*_{i}$ and $Q^*_{i}$ before and after secondary control, the superscription $(+)$ is added in Eq.(\ref{eq:gfsc}) to denote the values of $P^*_{i}$ and $Q^*_{i}$ after considering the secondary control. 

Similarly, if the DER at the bus $i$ is grid-following as shown in Fig. \ref{fig: primarycontrol}(b), the inverse droop control is defined as:
\begin{small}
\begin{equation}
\begin{aligned}
Inverse \quad Droop:
\left[ \begin{array}{l}{\bar{P}_{i_{}}}\\
\bar{Q}^{}_{i_{}}\end{array} \right]
=\left[ \begin{array}{l}
{-\frac{1}{\sigma_\omega}({\omega_{i_{}}}- \omega^*_{i_{}})}\\
-\frac{1}{\sigma_V}(V_{i_{}}- V^*_{i_{}})
\end{array} \right]
\label{eq:invdroop}
\end{aligned}
\end{equation}
\end{small}\noindent
where $\bar{P}_{i_{}}$ and $\bar{Q}_{i_{}}$ are the 
power generated by the inverse droop. The eventual real power $P_{i_{}}$ and reactive power $Q_{i_{}}$ sent to the grid-following control as references are: 
\begin{small}
\begin{equation}
\begin{aligned}
P_{i_{}}=\bar{P}_{i_{}} + P^*_{i_{}},\quad Q_{i_{}}=\bar{Q}_{i_{}}+ Q^*_{i_{}}
\label{eq:gfl}
\end{aligned}
\end{equation}
\end{small}\noindent
where $P^*_{i_{}}$ and $Q^*_{i_{}}$ are reference power guided by the secondary control. In a similar form to Eq. (\ref{eq:gfsc}), $P^*_{i_{}}$ and $Q^*_{i_{}}$ was updated in discrete time as: $P^{*(+)}_{i_{}}=P^{*}_{i_{}} + \Delta P^{*}_{i_{}}$ and $Q^{*(+)}_{i_{}}=Q^{*}_{i_{}} + \Delta Q^{*}_{i_{}}$.



Consequently, the droop for grid-forming control and the inverse droop for the grid-following control can be represented as: 
\begin{small}
\begin{subequations}
\begin{equation}
\begin{aligned}
\mbox{Droop:}
&\left[ \begin{matrix}{\omega_{i}-\omega^*_i}\\
{V_{i}-V^*_i}
\end{matrix} \right]
=\left[ \begin{matrix}
{-\sigma_\omega(P_i-P^{*(+)}_i)}\\
-\sigma_V(Q_i-Q^{*(+)}_i)
\end{matrix}\right]\\
& =\left[ 
\begin{matrix}
{-\sigma_\omega(P_i-P^*_i)}\\
{-\sigma_V(Q_i-Q^*_i)}
\end{matrix}
\right] + \left[\begin{matrix} \sigma_\omega & {} \\
{} & \sigma_V \end{matrix} \right]\bm{u}_i, \mbox{with }\bm{u}_i=\left[ \begin{matrix}
{\Delta P^*_i}\\
\Delta Q^*_i
\end{matrix}\right]
\label{eq:droopw}
\end{aligned}
\end{equation}
\begin{equation}
\begin{aligned}
\mbox{Inverse  Droop:}&
\left[\begin{matrix}{P_i-P^{*(+)}_i}\\
{Q_i-Q^{*(+)}_i}
\end{matrix} \right]
 =\left[ \begin{matrix}
{-\frac{1}{\sigma_\omega}({\omega_i}_{} - \omega^*_i)}\\
-\frac{1}{\sigma_V}(V_{i} - V^*_i)
\end{matrix} \right]\\
&\Rightarrow
\left[ \begin{matrix}{\omega_{i}-\omega^*_i}\\
{V_{i}-V^*_i}
\end{matrix} \right]=
\left[ \begin{matrix}
{-\sigma_\omega(P_i-P^*_i)}\\
-\sigma_V(Q_i-Q^*_i)
\end{matrix}\right]+\left[\begin{matrix} \sigma_\omega & \\
& \sigma_V \end{matrix} \right]\bm{u}_i
\label{eq:droopw2}
\end{aligned}
\end{equation}
\end{subequations}
\normalsize
\color{black} According to (\ref{eq:droopw})-(\ref{eq:droopw2})\color{black}, both the droop and the inverse droop take the same form: 
\small
\begin{small}
\begin{equation}
\begin{aligned}
\left[ \begin{matrix}{\dot{\theta}_i}\\
{\dot{V}_i}
\end{matrix} \right]=
\left[ \begin{matrix}{-\sigma_\omega(P_i-P^*_i)}\\
-\frac{\sigma_V}{\tau_V}(Q_i-Q^*_i)
\end{matrix}\right]
+
\left[\begin{matrix}
\sigma_\omega & \\
& \frac{\sigma_V}{\tau_V} \end{matrix}\right] \bm{u}_i \label{eq:droopandinversedroop}
\end{aligned}
\end{equation}
\end{small}\noindent
where
\begin{eqnarray} 
\dot\theta_i &=& \omega_i-\omega^*_i\\
\label{eq:thetaw}
P_i&=&\sum_{j=1}^{n}V_{i}V_{j}(G_{ij}\cos(\theta_i-\theta_j)+B_{ij}\sin(\theta_i-\theta_j))\\\label{eq:pf1}
Q_i&=&\sum_{j=1}^{n}V_{i}V_{j}(G_{ij}\cos(\theta_i-\theta_j)-B_{ij}\sin(\theta_i-\theta_j))\label{eq:pf2}
\end{eqnarray}
\end{small}\noindent
$\tau_V$ is the equivalent time constant of voltage magnitude dynamics due to the grid-forming or the grid-following control loops, which can be treated as a first-order inertia system when properly tuned; $\bm{u}$ denotes the external control inputs due to secondary control; $\theta$ is the voltage phasor angle; 
$j$ denotes the bus number; $G_{ij}$ and $B_{ij}$ represent the equivalent conductance and susceptance between bus $i$ and $j$. 

\color{black}As shown in Fig. \ref{fig: primarycontrol}, the droop generates the voltage reference for the grid-forming control system, and the inverse droop generates the power references for the grid-following control system. \color{black}
Note that the grid-forming control based on a PLL is adopted in this paper to mitigate negative \color{black} impacts on \color{black} systemwide stability \cite{GFM_PLL2019}. 
The dynamics of the grid-forming and grid-following loops are not presented in \color{black}detail \color{black} but will be considered in all simulations presented in Section \ref{sec:case studies}. Interested readers are referred to \cite{6200347,Wei2020} \color{black} for the detailed modeling.  

\emph{Uncertainty from Grid-Forming  Control:} 
As the local converter control 
is much faster than the secondary control, the voltage reference fed to grid-forming control in Fig. \ref{fig: primarycontrol}(a) is approximately equal to $\omega$ and $V$ assuming that the grid-forming control loop is well tuned. 
Nonetheless, when large disturbances (e.g., \color{black}MG transitions from the grid-connected mode to the islanded mode\color{black}, volatile generation and load) occur that \color{black}cause \color{black} large power perturbations, 
the nonlinearity driven by the system power flows 
(8)-(\ref{eq:pf2}) and \color{black} by \color{black} the control interaction between the droop module and the grid-forming control loops can emerge, \color{black}thus \color{black} leading to modeling uncertainty in Eq. (\ref{eq:droopandinversedroop}). 

\emph{Uncertainty from Grid-Following Control:} Likewise, the control interaction between the inverse-droop module and the grid-following control loops may emerge when there are system disturbances causing big perturbations to the angle and voltage. In addition, when there are large disturbances or measurement noises that make the grid voltage measurement distorted, the uncertainty due to the PLL can also directly introduce the modeling error to Eq. (\ref{eq:droopandinversedroop}) 
\cite{SRF-PLL}.




To describe the uncertainty from either the grid-forming or the grid-following control, Eq. (\ref{eq:droopandinversedroop}) can be modified as
\begin{small}
\begin{equation}
\begin{aligned}
\left[ \begin{matrix}{\dot{\theta}_i}\\
{\dot{V}_i}
\end{matrix} \right]=
\left[ \begin{matrix}{-\sigma_\omega(P_i-P^*_i)}\\
-\frac{\sigma_V}{\tau_V}(Q_i-Q^*)
\end{matrix}\right]
+
\left[\begin{matrix}
\sigma_\omega & \\
& \frac{\sigma_V}{\tau_V} \end{matrix}\right] \bm{u}_i +
\left[ \begin{matrix} f_\omega(P,Q,\theta, V)\\
f_V(P,Q,\theta,V)
\end{matrix} \right] \label{nldroopgf}
\end{aligned}
\end{equation}
\end{small}\noindent
where $f_\omega(.)$ and $f_V(.)$ are unknown nonlinear functions to describe the residual dynamics for the voltage phasor angle and magnitude. 

The aforementioned nonlinearity and uncertainty pose challenges to the conventional secondary control of MGs 
(e.g., model-based ones and PI) especially under large disturbances. To address these challenges, we propose a Koopman-inspired 
method that can help 
identify the system accurately and adaptively using data despite nonlinearity and uncertainty such that effective control can be designed. 
 
\section{Koopman-Inspired Identification and Control}
\subsection{Koopman Operator Theory}
Koopman operator theory \cite{Proctor2018} shows that a nonlinear dynamical system can be transformed into an infinite-dimensional linear system under a Koopman embedding mapping. The Koopman-enabled linear model is valid for global nonlinearity with the infinite-dimensional representation as opposed to traditional locally linearized small-signal models. However, in practice, one can consider finite-dimensional Koopman invariant subspaces where dominant dynamics can be described. Particularly, given a nonlinear dynamical system with external control $\bm{x}_{k+1}=F(\bm{x}_{k},\bm{u}_{k})$, where $\bm{x} \in \mathcal{M}$ and $\bm{u} \in \mathcal{U}$ with $\mathcal{M}$ and $\mathcal{U}$ being the manifolds of state and control input, we consider the Koopman embedding mapping $\bm{\Phi}$ from the two manifolds to a new Hilbert space $\bm\Phi: \mathcal{M} \times \mathcal{U} \to \mathcal{H}$, which lies within the span of the eigenfunctions $\varphi_{j}$. That is, $\bm{\Phi(x,u)}= \sum_{j=1}^{N_{\varphi}}\varphi_{j}(\bm{\bm{x},\bm{u}})\bm{v}_{j}$, where $\bm{\Phi}(\bm{x},\bm{u})=[\Phi_{1} (\bm{x,u}),\Phi_2(\bm{x,u}),…,\Phi_i(\bm{x,u}),…,\Phi_p(\bm{x,u})]^T$ is a set of Koopman observables, $\bm{v}_{j}$ are the vector-valued coefficients called Koopman modes. 
The Koopman operator $\mathcal{K}$, acting on the span of $\varphi_{j}$, advances the embeddings $\bm{\Phi(x,u)}$ linearly in the Hilbert space $\mathcal{H}$ as \cite{Proctor2018}:
\begin{small}
\begin{equation}
\begin{aligned}
\bm{\Phi}(\bm{x}_{k+1},\bm{u}_{k+1}) & =\mathcal{K}\Phi(\bm{x}_k,\bm{u}_k)\\
& = \mathcal{K}\sum_{j=1}^{N_\varphi}\varphi_j({\bm{x},\bm{u}})\bm{v}_j
= \sum_{j=1}^{N_\varphi}(\rho_{j}\varphi_j(\bm{x}_k,\bm{u}_k)\bm{v}_j)
 \end{aligned} \label{eq:edmd}
\end{equation}\end{small}\noindent
where $\rho_{j}$ are the eigenvalues satisfying $\mathcal{K} \varphi_j(\bm{x,u})=\rho_j \varphi_j(\bm{x,u})$. To be consistent with the linear form of control inputs in Eq. (\ref{nldroopgf}), we assume that $\Phi_i(\bm{x,u})=g_{i}(\bm{x})+l_i(\bm{u})$ where $g_{i}(\bm{x})$ is a nonlinear observable \color{black} function and  $l_i(\bm{u})$ is linear with $l_i(\bm{0})=0$ 
\cite{KORDALinear2018}. In addition, we assume $\Phi_{i}(\bm{x}_{k+1},\bm{0}) = \mathcal{K} \Phi_{i}(\bm{x}_k,\bm{u}_k)$ for all $k$. Then, $g_{i}(\bm{x}_{k+1})+l_i(\bm{0})=\mathcal{K} g_{i}(\bm{x}_k)+\mathcal{K} l_{i}(\bm{u}_k)  \Rightarrow g_{i}(\bm{x}_{k+1})=\mathcal{K} g_{i}(\bm{x}_k)+\mathcal{K} l_i(\bm{u}_k)$. This assumption means that the Koopman operator is only attempting to propagate the observable functions at the current state $\bm{x}_k$ and inputs $\bm{u}_k$ to the future observable functions on the state $\bm{x}_{k+1}$ but not on future inputs $\bm{u}_{k+1}$ (i.e., $[\bm{\Delta P}^*,\bm{\Delta Q}^{*} ]^T$ are not state-dependent) \cite{Proctor2018}. Let us define $\bm{z}:=\bm{g}(\bm{x})=[g_1(\bm{x}),g_2(\bm{x}),…,g_i(\bm{x}),…g_p(\bm{x})]^T$. Then we have an approximation of Eq. (\ref{nldroopgf}) in a form of extended dynamic mode decomposition with control (EDMDc) \cite{KORDALinear2018} as below 
\begin{small}
\begin{subequations}
\begin{equation}
(Process \quad  model) \quad \bm{z}_{k+1} = \bm{A}\bm{z}_k+\bm{B}\bm{u}_k+\bm{\delta}_k
\label{eq:process}\end{equation}
\begin{equation}
(Observation \quad  model) \qquad  \bm{y}_k = \bm{C}\bm{z}_k+\bm{e}_k
\label{eq:observation}
\end{equation}\end{subequations}\end{small}\noindent
where $\bm{y}_k$ are the outputs of the Koopman state space model.
We define $\bm{y}_k =[d\theta_k,dV_k]^T= [\theta_k-\theta_{L}^*,V_k-V_L^*]^T $ in Eq. (\ref{eq:observation}) as the PMU-measured phasor angle and voltage magnitude deviations from the local operation points $[\theta_{L}^*,V_L^*]^T$ that are 
the first data sample from a window of collected PMU data. 
$\bm{A}$ and $\bm{B}$ are the state transition matrix and control matrix, satisfying that $\bm{Az}_k=\mathcal{K} \bm{g}(\bm{x}_k)$ and $\bm{Bu}_k=\mathcal{K} \bm{l}(\bm{u}_k)$. $\bm{\delta}_{k}$ is the Koopman modeling error associated with the EDMDc approximation. $\bm{e}_k$  is the observation model error. Given proper Koopman observables $\bm{z}$, the Koopman state space model (\ref{eq:process})-(\ref{eq:observation}) can describe large signal-driven nonlinear dynamics. That being said, under the Koopman embedding $\bm{z}=\bm{g}(\bm{x})$, the nonlinear dynamical system (\ref{nldroopgf}) can be represented by the linear dynamical system (\ref{eq:process})-(\ref{eq:observation}) that is valid under both small and large perturbations. There are three consecutive tasks to use this model for control: determination of Koopman observables, online identification of the Koopman state space model, and implementation of linear control (illustrated in Sections 3.2, 3.3 and 3.4, respectively). 

\subsection{Koopman Observables for MG Secondary Control}
\color{black} The selection of Koopman observables is important for realizing accurate modeling. The observables can be selected either empirically \cite{KORDA2018297, Netto2021, Gong2022} or with the help of machine learning techniques \cite{Yeung2019Koopman,Shi2022,Han2020}, while it remains an open question to obtain the best possible observables. In this paper, we selected the Koopman observables based on our experience and domain knowledge of power systems and microgrids. \color{black}According to Eq. (\ref{eq:droopandinversedroop})-Eq. (\ref{eq:pf2}), sinusoidal-driven interaction dynamics may emerge when subject to large perturbations and low inertia (i.e., the general solution for the droop-control differential equations contains trigonometric patterns). Inspired by this, we include the functions $\sin\theta$ and $\cos\theta$ into the Koopman embedding to describe such underlying dynamics, which were shown effective to describe interaction transients of power grids \cite{KORDA2018297}. Thus, let us define the MG original states $\bm{x}_k=[\theta_k, V_k]^T$ and the Koopman real-valued observables $\bm{z}_k=\bm{g}(\bm{x}_k)$ as:
\begin{small}
\begin{equation}
\begin{aligned}
\bm{z}_k
=\bm{g}(\bm{x}_k)
=[\Delta \bm{V}_k, \sin\bm{\theta}_k-\sin(\bm{\theta}_{L,k}^*), \cos\bm{\theta}_k-\cos(\bm{\theta}_{L,k}^*), \Delta\bm{\omega}_k]^T
 \end{aligned} \label{eq:bases}
\end{equation}
\end{small}\noindent
where $\bm{\Delta V}$ and $\bm{\Delta\omega}$ are voltage and angular frequency deviations from the nominal values. $\bm{\theta}_{L,k}^*$ represents the approximate underlying operation point of voltage phasor angle at time step $k$.  
The Koopman observables $\bm{z}$  constitute the Koopman state space in the form of Eq. (\ref{eq:process})-Eq. (\ref{eq:observation}), where the parameter matrices $\bm{A}$, $\bm{B}$ and $\bm{C}$ are to be determined by an advanced system identification method online as described in the next section. 
\subsection{Online Identification: A Koopman-Inspired Enhanced OKID Algorithm}
Considering the Koopman-based linear dynamical system model (\ref{eq:process})-(\ref{eq:observation}), we propose 
an observer Kalman filter identification (OKID)–based optimization algorithm to optimally identify the MG Koopman state space model (i.e., the matrix parameters $\bm{A}$,$\bm{B}$ and $\bm{C}$).

\textit{\textbf{
The OKID Algorithm}}. \color{black} Belonging to the category of closed-loop subspace methods, the conventional OKID algorithm is commonly used to identify linear systems \cite{QIN20061502}. \color{black}It is free of the bias problem that most typical closed-loop subspace methods have \cite{QIN20061502}, and has been applied in many areas such as aircraft control and autonomous underwater vehicles \cite{alenany_modified_2019}. 
In OKID, \color{black}the \color{black} impulse response of the system is estimated in a least-squares fashion with data. Then, a state space model of the system is obtained with the eigensystem realization algorithm (ERA). Specifically, let $\bm{Y}$ 
and $\bm{U}$ represent the matrix stacking the time series data of the outputs $\bm{y}$
and the control inputs $\bm{u}$ in a matrix form. Let $\bm{Y}_i$ and $\bm{U}_i$ represent the observation outputs and the control inputs at the $i^{th}$ time step in the data matrix, and consider the length of the sliding window is $N$. \color{black} By observing Eq. (\ref{eq:process})-Eq. (\ref{eq:observation}) and assuming zero initial conditions, 
$\bm{y}_k$ can be expressed with iterations in a form of 
$\bm{y}_k=\bm{C}\bm{z}_k=\bm{C}(\bm{A}\bm{z}_{k-1}+\bm{B}\bm{u}_{k-1})=\bm{C}(\bm{A}(\bm{A}\bm{z}_{k-2}+\bm{B}\bm{u}_{k-2})+\bm{B}\bm{u}_{k-1})= \bm{CA^{k-1}Bu}_{0} + \bm{CA^{k-2}Bu}_{1}+... \bm{CBu}_{k-1}=\sum_{i=0}^{k-1}\bm{C}\bm{A}^{k-i-1}\bm{B}\bm{u}_{i}$, whereby we obtain 

\begin{small}
\begin{equation}
\begin{aligned}
\bm{Y}
=\begin{bmatrix}
\bm{CB} & \cdots & \bm{CA^{N-1}B}
\end{bmatrix} 
\begin{bmatrix}
\bm{U}_0 & \bm{U}_1  & ...& \bm{U}_{N-1}\\
\bm{0} & \bm{U}_0 & ...&\bm{U}_{N-2}\\
\vdots & \vdots & \ddots & \vdots \\
\dots & \dots & \dots &\bm{U}_{0}
\end{bmatrix}\label{eq:diagnoalMatrix}
\end{aligned}\end{equation} 
\end{small}\noindent

Let $\bm{h}$ denote the 
impulse response of the Koopman state space model (\ref{eq:process})-(\ref{eq:observation}) in the sliding window of size $N$ (from $k=1$ to $k=N$) 
with zero initial conditions ($\bm{x}_0=0$) and impulse inputs ($\bm{u}_0=1$ and $\bm{u}_k=0$ when $k>0$), we have 
\begin{small}
\begin{equation}
\begin{aligned}
\bm{h}&=\begin{bmatrix}
\bm{h}_1  & \bm{h}_2 & ...& \bm{h}_N
\end{bmatrix}
&=\begin{bmatrix}
\bm{CB} & \bm{CAB} & ...& \bm{CA^{N-1}B}
\end{bmatrix}\\
 \end{aligned} \label{eq:impulse}
\end{equation}
\end{small}\noindent
Then according to Eq. (\ref{eq:diagnoalMatrix})- (\ref{eq:impulse}) and with the knowledge of the observation matrix $\bm{Y}$ and the control input matrix $\bm{U}$, one can estimate the impulse response in a least-squares fashion 
\begin{small}
\begin{equation}
\begin{aligned}
\begin{bmatrix}
\bm{h}_1  & \bm{h}_2 & ...& \bm{h}_N
\end{bmatrix}=\bm{Y}\begin{bmatrix}
\bm{U}_0 & \bm{U}_1  & ...& \bm{U}_{N-1}\\
\bm{0} & \bm{U}_0 & ...&\bm{U}_{N-2}\\
\vdots & \vdots & \ddots & \vdots \\
\bm{0} & \bm{0} & \dots &\bm{U}_{0}
\end{bmatrix}^{\dagger}
 \end{aligned} \label{eq:OKID}
\end{equation}
\end{small}\noindent
where the operator $\dagger$ represents the Moore-Penrose pseudo-inverse. \color{black} Note that the noise is not optimally filtered by the least-squares inverse as presented in Eq. (\ref{eq:OKID}). To address the issue, the conventional OKID can be designed based on an optimal observer system whereby optimal system parameters can be identified. For simplicity, we refer readers to \cite{brunton2022data} (Pages 340-343) for detailed explanation and implementation.

Next, with the obtained impulse response, the Hankel matrix $\bm{H}$ and the next-step Hankel matrix $\bm{H^{'}}$ can be written as follows: \color{black} 
\begin{small}
\begin{equation}
\begin{aligned}
\bm{H}=\begin{bmatrix}
\bm{h}_1 & \bm{h}_2  & ...& \bm{h}_{N}\\
\bm{0} & \bm{h}_1 & ...&\bm{h}_{N-1}\\
\vdots & \vdots & \ddots & \vdots \\
\bm{0} & \bm{0} & \dots &\bm{h}_{1}
\end{bmatrix}, 
\bm{H^{'}}=\begin{bmatrix}
\bm{h}_2 & \bm{h}_3  & ...& \bm{h}_{N+1}\\
\bm{0} & \bm{h}_2 & ...&\bm{h}_{N}\\
\vdots & \vdots & \ddots & \vdots \\
\bm{0} & \bm{0} & \dots &\bm{h}_{2}
\end{bmatrix}
 \end{aligned} \label{eq:Hankels}
\end{equation}
\end{small}\noindent
The Hankel matrix $\bm{H}$ could be truncated with Singular Value Decomposition (SVD):
\begin{small}
\begin{equation}
\begin{aligned}
\bm{H}= \bm{\mathcal U} \bm{\Sigma} \bm{\mathcal V}^T=[\bm{\widetilde{\mathcal U}}, \bm{\mathcal{U}_{tr}}]\begin{bmatrix}
\bm{\widetilde\Sigma} & \bm{0} \\
\bm{0} & \bm{\Sigma_{tr}} \end{bmatrix}
\begin{bmatrix}
\bm{\widetilde{\mathcal V}}^T \\
\bm{\mathcal{V}_{tr}}^T \end{bmatrix}
\approx  \bm{\mathcal {\widetilde {U}}} \bm{\widetilde\Sigma} \bm{\mathcal{\widetilde V}}^T
 \end{aligned} \label{eq:SVD}
\end{equation}
\end{small}\noindent
 Let 
 \begin{equation}
\color{black} \bm{\mathcal{O}} =[\bm{C},\bm{CA},\bm{CA}^2,...,\bm{CA}^{N-1}]^T\label{eq:O}
\end{equation}
 be the observability matrix, and 
 \begin{equation}
\color{black} \bm{\mathcal{C}} =[\bm{B},\bm{AB},\bm{A}^2\bm{B},...,\bm{A}^{N-1}\bm{B}]\label{eq:C}
 \end{equation}
 be the controllability matrix. Then, by observing Eq. (\ref{eq:impulse}) and Eq. (\ref{eq:Hankels}), we have 
 \begin{equation}
 \bm{H}=\bm{\mathcal{O}\mathcal{C}},\quad \bm{H^{'}}=\bm{\mathcal{O}\bm{A}\mathcal{C}} \label{eq:H}
\end{equation}
Furthermore, considering Eq. (\ref{eq:SVD}),  
 we can assume that $\bm{\mathcal{O}}=\bm{\widetilde{\mathcal U}}\bm{\widetilde \Sigma}^{\gamma}$ and $\bm{\mathcal{C}}=\bm{\widetilde \Sigma}^{1-\gamma}\bm{\widetilde{\mathcal V}}^T$, where $\gamma$ is an arbitrary real value. 

\emph{Conventional OKID algorithm}. For the conventional OKID algorithm, ERA is thereafter used to identify the matrix $\bm{A}$ and $\bm{B}$, with $\gamma$ set to a constant $\frac{1}{2}$ for a special balanced realization.
That is, one can assume $\bm{\mathcal{O}}=\bm{\widetilde{\mathcal U}}\bm{\widetilde \Sigma}^{\frac{1}{2}}$ and $\bm{\mathcal{C}}=\bm{\widetilde \Sigma}^{\frac{1}{2}}\bm{\widetilde{\mathcal V}}^T$, whereby a state space model with balanced Grammians is realized (i.e., the same degree of controllability and observability) that agrees with the control input and the observation data. \color{black} As such, with $\gamma=\frac{1}{2}$ and by Eq. (\ref{eq:O}) - Eq. (\ref{eq:H}), the matrices $\bm{A}$ and $\bm{B}$ can be identified by the \textbf{conventional OKID} as follows \cite{brunton2022data}:
\color{black}
\begin{small}
\begin{subequations}
\begin{equation}
\bm{\widetilde{A}}=\bm{\widetilde \Sigma}^{-\frac{1}{2}}\bm{\widetilde{\mathcal U}}^{T}\bm{H}^{'}\bm{\widetilde{\mathcal V}}\bm{\widetilde \Sigma}^{-\frac{1}{2}}
\end{equation}
\begin{equation}
\bm{\widetilde{B}}=\mathfrak{\bm{C}}_{N_S \times N_U}=\left[\bm{\widetilde \Sigma}^{\frac{1}{2}}\bm{\widetilde{\mathcal V}}^T\right]_{N_S \times N_U}
\end{equation}\label{eqs:cOKID_AB}\end{subequations}
\end{small}\noindent
where the operator $\begin{bmatrix}.\end{bmatrix}_{N_S \times N_U}$ represents the first $N_S$ rows and the first $N_U$ columns of the matrix in the bracket; $N_S$ is the dimension of Koopman embedding space and $N_U$ is the dimension of control inputs. 

\begin{figure}[!tb]
\centering
  \includegraphics[width=1\linewidth]{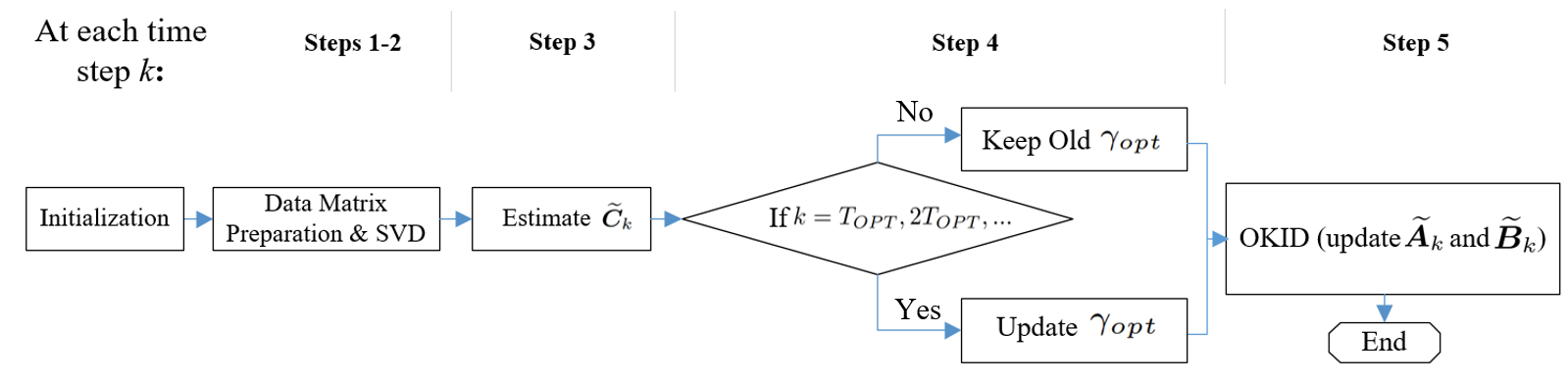}
  \caption{Algorithm flowchart of the proposed Koopman-inspired enhanced OKID}
  \label{FlowchartOKID}
\end{figure}

In this paper, to better identify the Koopman-based process dynamics, we propose a Koopman-inspired algorithm to find an optimal $\gamma$ rather than assuming $\gamma = \frac{1}{2}$ as in the conventional OKID. \color{black} Consider a general form with $\gamma$ unfixed 
\begin{small}
\begin{subequations}
\begin{equation}
\begin{aligned}
\bm{\widetilde{\mathcal U}}\bm{\widetilde \Sigma}^{\gamma}
& =\bm{\mathcal{O}}=\bm{[C,CA,CA^2,...,CA^{N-1}]}^T\\
& =
 \color{black}\bm{I}_{N\times N}\otimes\bm{C}\cdot
 \bm{[I,A,A^2,...,A^{N-1}]^T}
 \end{aligned}\label{eq:2EstimateC}
\end{equation}
\begin{equation}
\begin{aligned}
\bm{\widetilde \Sigma}^{1-\gamma} \bm{\widetilde{\mathcal V}}^T
& =\bm{\mathcal{C}}
=\bm{[B,AB,A^2B,...,A^{N-1}B]}\\
& = \bm{[I,A,A^2,...,A^{N-1}]}
\color{black}(\bm{I}_{N\times N}\otimes \bm{B})
 \end{aligned}\label{eq:2EstimateB}
\end{equation}
\end{subequations}
\end{small}\noindent
where $\bm{I}_{N \times N}$ is the identity matrix with the dimension $N \times N$, and $\otimes$ denotes the Kronecker product which is
\begin{equation} 
\color{black}\bm{I}_{N\times N}\otimes \bm{C}=
\begin{bmatrix}
\bm{C} &  & \\
 & \ddots &  \\
&  & \bm{C} \end{bmatrix}, \quad
\color{black}\bm{I}_{N\times N}\otimes \bm{B}=
\begin{bmatrix}
\bm{B} &  & \\
 & \ddots &  \\
&  & \bm{B} \end{bmatrix}
\end{equation}
Then
\begin{small}
\begin{subequations}
\begin{equation}
\begin{aligned}
 (\color{black}\bm{I}_{N\times N}\otimes\bm{C})^{\dagger}
 \bm{\widetilde{\mathcal U}}\bm{\widetilde \Sigma}^{\gamma}
 =\bm{[I,A,A^2,...,A^{N-1}]^T}
 \end{aligned} \label{eq:ASeries1}
\end{equation}
\begin{equation}
\begin{aligned}
\bm{\widetilde \Sigma}^{1-\gamma} \bm{\widetilde{\mathcal V}}^T
 \color{black}(\bm{I}_{N\times N}\otimes \bm{B})^{\dagger}
 = \bm{[I,A,A^2,...,A^{N-1}]}
 \end{aligned} \label{eq:ASeries2}
\end{equation}
\end{subequations}
\end{small}\noindent
By observing Eq. (\ref{eq:ASeries1}) and Eq. (\ref{eq:ASeries2}), we have
\begin{small}
\begin{equation}
\begin{aligned}
&\bm{\widetilde \Sigma}^{\gamma}\bm{\widetilde{\mathcal U}}^T
 \color{black}((\bm{I}_{N\times N}\otimes\bm{C})^{\dagger}
 )^{T}
 =\bm{\widetilde \Sigma}^{1-\gamma} \bm{\widetilde{\mathcal V}}^{T}
\color{black}(\bm{I}_{N\times N}\otimes\bm{B})^{\dagger}\\
&\Rightarrow\bm{\widetilde \Sigma}^{2\gamma-1}\bm{\widetilde{\mathcal U}}^T
\begin{pmatrix}
 \color{black}(\bm{I}_{N\times N}\otimes \bm{C})^{\dagger} \end{pmatrix}^{T}
 =\bm{\widetilde{\mathcal V}}^{T}
\color{black}(\bm{I}_{N\times N}\otimes \bm{B})^{\dagger}
 \end{aligned} \label{eq:OKID-ERA}
\end{equation}
\end{small}\noindent
Treating Eq. (\ref{eq:OKID-ERA}) 
as a soft constraint for the parameter $\gamma$, one can formulate a quadratic optimization problem to solve the optimal parameter $\gamma_{opt}$:
\begin{small}
\begin{equation}
\begin{aligned}
\gamma_{opt}=arg 
\min_{\gamma}
\|
\bm{\widetilde \Sigma}^{2\gamma-1}\bm{\widetilde{\mathcal U}}^T
\begin{pmatrix}

\color{black}(\bm{I}_{N \times N} \otimes \bm{C})^{\dagger}\end{pmatrix}^{T} -\bm{\widetilde{\mathcal V}}^{T}
\color{black}(\bm{I}_{N\times N} \otimes \bm{B}) \|_{F}
\end{aligned}
\nonumber
\end{equation}\label{eq:OKID-ERA-Implement1}
\begin{equation} \mbox{subject  to:} \qquad 0\leq \gamma \leq 1
\label{eq:OKID-ERA-Implement2}\end{equation}
\end{small}\noindent
where $\|.\|_F$ represents the Frobenius norm of a matrix. 
The inequality $0\leq \gamma \leq 1$
is added to constrain problem complexity. The novel OKID-based algorithm for parameter estimation is summarized below. The flowchart of the algorithm is also presented in Fig. \ref{FlowchartOKID}.

\noindent\textit{\textbf{ 
The Proposed Online Koompan-Inspired Enhanced OKID Algorithm}} 
\noindent 
\textbf{Algorithm Initialization.} 
Initialize $\gamma_{opt}=\gamma_{opt,0}$, the smoothing factor $\eta$, and the time step $T_{OPT}$ between two updates of $\gamma_{opt}$. 
The selection of these parameters will be discussed in \textit{Remarks} after  the presentation of the algorithm. 


At each time step  of identification and the secondary control, i.e., for $k=1, 2,...$,  conduct \textbf{Step 1} -\textbf{Step 5}.

\noindent
\textbf{Step 1: Data preparation.} Collect the last $N$ data samples from microPMUs to obtain the data matrices of phasor angle $\bm{\Theta}$, voltage deviation $\bm{\Delta V}$ and angular frequency deviation $\bm{\Delta \Omega}$ from the nominal values. 
Collect control input data $\bm{U}$ from the secondary controller. For example, the phasor angle $\bm{\Theta}$ is stacked in a form of
\begin{small}
\begin{equation}
\begin{aligned}
\bm{\Theta} &=\begin{bmatrix}
| & | & |\\
\bm{\Theta}_1 & ... & \bm{\Theta}_{N}\\
| & | & | \end{bmatrix}
\end{aligned}\label{eq:data}
\end{equation}
\end{small}\noindent
$\bm{\Delta V}$, $\bm{\Delta \Omega}$ and $\bm{U}$ are formed in the same way. \color{black} The approximated operation points of voltage phasor angles and magnitudes $\bm{\Theta_L^*}$ and $\bm{V_L^*}$ are defined as the first data sample from a window of collected PMU data, prepared in a matrix form as follows:
\color{black}
\begin{small}
\begin{equation}
\begin{aligned}
\bm{\Theta_L^{*}} &=\begin{bmatrix}
| & | & |\\
\bm{\Theta}^{}_1 & ... & \bm{\Theta}_{1}^{}\\
| & | & | \end{bmatrix},\quad
\bm{V_L^{*}}=\begin{bmatrix}
| & | & |\\
\bm{V}^{}_1 & ... & \bm{V}_{1}^{}\\
| & | & | \end{bmatrix}
\end{aligned}\label{eq:ldata}
\end{equation}
\end{small}\noindent
\color{black}
Prepare the data matrices for  $\bm{y}$ and $\bm{z}$ as follows:  $\bm{Y=[\Theta-\Theta^*_L,V-V^*_L]^T}$, and
$\bm{Z=[\Delta V,\sin(\Theta)-\sin(\Theta^*_L),\cos(\Theta)-\cos(\Theta^*_L), \Delta\Omega]^T}$.

\noindent \textbf{Step 2: Hankel matrix preparation and SVD}. 
Estimate the impulse response and prepare the Hankel matrices according to Eq. (\ref{eq:OKID})-Eq. (\ref{eq:Hankels}). \color{black} Conduct the SVD on the obtained Hankel matrix $\bm{H}\approx \bm{\mathcal {\widetilde {U}}} \bm{\widetilde\Sigma} \bm{\mathcal{\widetilde V}}^T$. 

\noindent \textbf{Step 3:  Estimation of $\bm{C}$.} 
\color{black}Ignoring the error term in Eq. (\ref{eq:observation}), we have $\bm{Y} = \bm{C}\bm{Z}$. Thus, one can estimate the observation matrix $\bm{C}$ at each time step $k$ in a least-squares fashion by multiplying the pseudo-inverse on both sides of the equation, which is 
\begin{small}
\begin{equation}
\begin{aligned}
\bm{\widetilde{C}}_k &= \bm{Y}\bm{Z}^{\dagger}
\end{aligned}\label{eq:Cest}
\end{equation}
\end{small}\noindent
\color{black}
\noindent \textbf{Step 4: Optimization for $\gamma_{opt}$.} 
Check if the run time of optimization between the last update of $\gamma_{opt}$ is larger than $T_{OPT}$. If no, $\gamma_{opt,k}=\gamma_{opt,k-1}$, go to \textbf{Step 5}; otherwise, solve the optimization problem in Eq. (\ref{eq:OKID-ERA-Implement2}) for $\gamma_{opt}^-$. 
To do so, by Eq. (\ref{eqs:cOKID_AB}b), 
replace $\bm{B}$ with $\begin{bmatrix} \bm{\widetilde \Sigma}^{1-\gamma} \bm{\widetilde{\mathcal V}}^T \end{bmatrix}_{N_S \times N_U}$ and replace $\bm{C}$ with $\bm{\widetilde{C}}_k$ from \textbf{Step 3} 
in 
Eq. (\ref{eq:OKID-ERA-Implement2}). 
Then, adaptively update $\gamma_{opt}$ by
\begin{small}
\begin{equation}
\begin{aligned}
\gamma_{opt,k} =  
&\eta\gamma_{opt}^{-} + (1-\eta)\gamma_{opt,k-1}, & {\mbox{for} \quad k = T_{OPT},2T_{OPT},3T_{OPT},...} 
\end{aligned}
\end{equation}\end{small}\noindent
\color{black} where $\gamma_{opt}^{-}$ is the optimal value of the realization parameter $\gamma$ according to Eq. (\ref{eq:OKID-ERA-Implement2}). That is, once $\gamma_{opt}^{-}$ is updated,  we update $\gamma_{opt}$ with the weighted sum of the old $\gamma_{opt}$ at last time step and the updated value $\gamma_{opt}^{-}$. $\eta$ is the weight to smooth online learning. The role of $\eta$ is to smooth the online learning of $\gamma$. As the small piece of online data used for identification is characterized by stochasticity, the smoothing factor $\eta$ can mitigate aggressive change to make the learning process more reliable. This is so because the estimation is equivalent to the Robbins–Monro form \cite{haykin2009neural}, which is $\gamma_{opt,k}=\eta\gamma_{opt}^{-} + (1-\eta)\gamma_{opt,k-1} = \gamma_{opt,k-1} +\eta(\gamma_{opt}^{-} -\gamma_{opt,k-1})$.
The larger the value of $\gamma$ is, the smoother the learning process tends to be, whereas the adaptiveness of learning is compromised.
\color{black}

\noindent \textbf{Step 5: Estimation of $\bm{A}$ and $\bm{B}$}. 
By Eq. (\ref{eqs:cOKID_AB}a)-(\ref{eqs:cOKID_AB}b)
\begin{small}
\begin{equation}
\begin{aligned}
\bm{\widetilde{A}}_k = &\left\{
\begin{array}{rc}
\eta\bm{\widetilde \Sigma}^{-\gamma_{opt,k}}\bm{\widetilde{\mathcal U}}^{T}\bm{H}^{'}\bm{\widetilde{\mathcal V}}\bm{\widetilde \Sigma}^{\gamma_{opt,k}-1}+ (1-\eta)\bm{\widetilde{A}}_{k-1} & {\mbox{if } k \geq 1} \\
\bm{\widetilde \Sigma}^{-\gamma_{opt,k}}\bm{\widetilde{\mathcal U}}^{T}\bm{H}^{'}\bm{\widetilde{\mathcal V}}\bm{\widetilde \Sigma}^{\gamma_{opt,k}-1}\qquad\qquad & \mbox{if }{k=0} \\
\end{array}\right.
\end{aligned}\label{OKID_A}\end{equation}\end{small}
\begin{small}
\begin{equation}
\begin{aligned}
\bm{\widetilde{B}}_k = &\left\{
\begin{array}{rc}
\eta\left[\bm{\widetilde \Sigma}^{1-\gamma_{opt,k}}\bm{\widetilde{\mathcal V}}^T\right]_{N_S \times N_U}+(1-\eta)\bm{\widetilde{B}}_{k-1} & {\mbox{if } k \geq 1} \\
\left[\bm{\widetilde \Sigma}^{1-\gamma_{opt,k}}\bm{\widetilde{\mathcal V}}^T\right]_{N_S \times N_U} \qquad\qquad\qquad & \mbox{if }{k=0} \\
\end{array}\right.\\
\end{aligned}\label{OKID_B}\end{equation}\end{small}\noindent
\color{black}After implementing the identification algorithm, the identified Koopman state space model at the time step $k$ is obtained as:
\begin{small}
\begin{subequations}
\begin{equation}
\begin{aligned}
\bm{z}_{k+1} = \bm{\widetilde A}_k\bm{z}_k+\bm{\widetilde B}_k\bm{u}_k	
\end{aligned}\label{eq:idKoopman1}
\end{equation} 
\begin{equation}
\begin{aligned}
\bm{y}_k = \bm{\widetilde C}_k\bm{z}_k
 \end{aligned} \label{eq:idKoopman2}
\end{equation}
\end{subequations}
\end{small}

\color{black} Compared to the traditional EDMDc used in power systems \cite{KORDA2018297}, the proposed Koopman-inspired OKID can use the observation data $\bm{y}$ as in Eq. (\ref{eq:observation}) to help learn the Koopman state space model in Eq. (\ref{eq:process}), while the traditional EDMDc only estimates the Koopman state space in Eq. (\ref{eq:process}) in a least-squares fashion without the incorporation of observation data. The fusion of the information from the observation data provides extra opportunities to enhance the modeling efficacy.\color{black}\\

\noindent{\emph{Remarks}} 
\begin{itemize}
    \item $\gamma_{opt}$: 
    in this paper, $\gamma_{opt,0}=\frac{1}{2}$. Thus the enhanced OKID is initially equivalent to the conventional one while it gradually learns the optimized value for $\gamma_{opt}$ with the online OKID and the periodically enabled optimization in Eq. (\ref{eq:OKID-ERA-Implement2}). 

    \item The smoothing factor $\eta$: it is used to \color{black}weigh \color{black} the past estimations and the latest one, and set to $\frac{1}{N}$ in this paper with the assumption that all estimations have the same weight independent on the time of occurrence. A larger $\eta$ means the estimation put more \color{black} weight \color{black} on the newest data, and vice versa. \color{black}

    \item The time step $T_{OPT}$ for updating $\gamma_{opt}$: it is set to 0.6s, which is longer than the run time of the proposed Koopman-inspired enhanced OKID and the time step of secondary control (30ms) as detailed in Section \ref{sec:case studies}. A small $T_{OPT}$ is favorable as a fast update of $\gamma$ to 
    compensate for \color{black} the \color{black} uncertainty of the Koopman process model (\ref{eq:process}), while it should be longer than the run time of the optimization (\ref{eq:OKID-ERA-Implement2}) to ensure the feasibility of online implementation. 
\end{itemize}
\color{black}

\subsection{The Linear Control Based on the Koopman-Inspired Enhanced OKID} 
After obtaining the identified model (\ref{eq:idKoopman1}) - (\ref{eq:idKoopman2}), a discrete-time linear quadratic regulator (LQR) is applied at each time step of secondary control,  aiming to reduce the voltage and frequency deviations by minimizing the cost
\begin{small}
\begin{equation}
\begin{aligned}
& \bm{J}(\bm{u})= \sum_{k=0}^{\infty}\bm{z}_{k}^{T}\bm{\mathcal{Q}}\bm{z}_k+\bm{u}_k^T\bm{\mathcal{R}}\bm{u}_k,\\
& \mbox{subject  to} \quad \bm{z}_{k+1} = \bm{\widetilde A}\bm{z}_k+\bm{\widetilde B}\bm{u}_k	
 \end{aligned} \label{eq:LQRcost}
\end{equation}
\end{small}\noindent
where $\bm{\mathcal{Q}}$ and $\bm{\mathcal{R}}$ are cost matrices defined as:
\begin{small}
\begin{equation}
\bm{\mathcal{Q}}=\begin{bmatrix}
\bm{ q}_V & & & \\
& \bm{ q}_{\sin \theta} &  &\\
& & \bm{ q}_{\cos \theta} & \\
& & & \bm{ q}_{\omega} \end{bmatrix},\quad
\bm{\mathcal{R}}=\begin{bmatrix}
\bm{{r}}_P &  \\
& \bm{{r}}_Q \end{bmatrix}
\label{eq:LQRcost}
\end{equation}
\end{small}\noindent
where $\bm{q_{V}}$, $\bm{q_{\sin\theta}}$, $\bm{q_{\cos\theta}}$ and $\bm{q}_{\omega}$ are cost submatrices for the Koopman observables presented in (\ref{eq:bases}). $\bm{r_P}$ and $\bm{r_Q}$ are cost submatrices for the control signals $\Delta P^*$ and $\Delta Q^*$. \color{black}They are basically selected empirically in this paper based on which factor is treated to be more important. \color{black}The optimal control input can be obtained by:
\begin{equation}
\begin{small}
\begin{aligned}
\bm{u}_k = &\left\{
\begin{array}{rcl}
\bm{U_{LB}} & {} & {\bm{u}_k < \bm{U_{LB}} }\\
-\bm{K}\bm{z}_{k} &  & { \bm{U_{LB}} \leqslant \bm{u}_k \leqslant \bm{U_{UB}} }\\
\bm{U_{UB}} &  & {\bm{u}_k > \bm{U_{UB}}} \\
\end{array}\right.\\
& \mbox{with }\quad \bm{K} = (\bm{\widetilde{B}}^T\bm{S}\bm{\widetilde{B}}+\bm{\mathcal{R}})^{-1}\bm{\widetilde{B}}^T\bm{S}\bm{\widetilde{A}}
\end{aligned}
\end{small}
\end{equation}\label{LQR}\noindent
where $\bm{K}$ is the control gain matrix. $\bm{S}$ is the solution of Riccati equation \cite{LQRformula2008}. $\bm{U_{UB}}$ and $\bm{U_{LB}}$  are the upper and lower saturation limits that can bound the uncertainty introduced by control inputs. \color{black} The bounds are user-defined values, which are determined empirically in the paper. Usually, large bounds can lead to faster response whereas the uncertainty introduced through control input channels could be increased to an unmanageable level that degrades the dynamic control performance or even stability. On the other hand, the bounds cannot be set to too small values, otherwise, the response could be slow and the capability of the controller cannot be fully taken use of. \color{black} 

\begin{figure}[h]
\centering
  \includegraphics[width=1\linewidth]{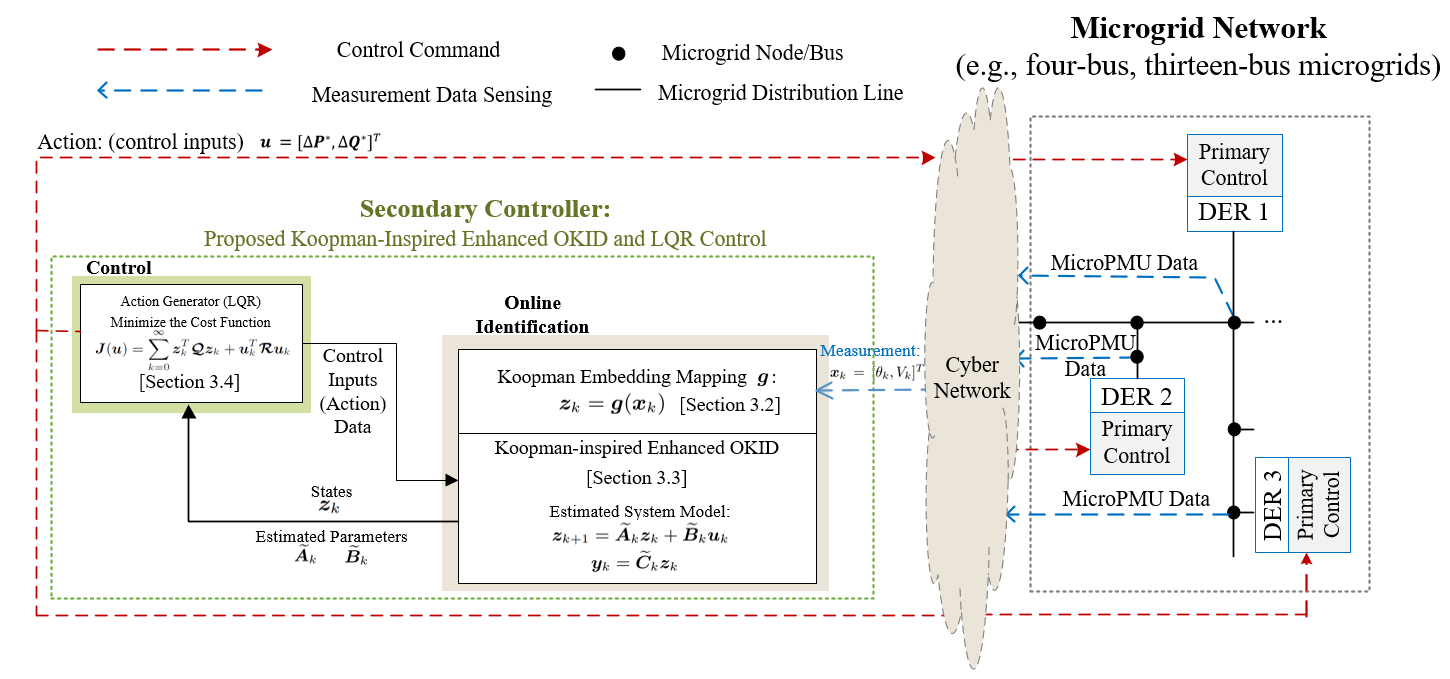}
  \caption{\color{black} Online structure of the proposed Koopman-inspired enhanced OKID and control}
  \label{Flowchart}
\end{figure}

\color{black} The Koopman-inspired enhanced OKID illustrated in Section 3.3 and the LQR illustrated in Section 3.4 can be respectively applied to the identification block and the control algorithm block of secondary control in Fig. \ref{MGarch}. Specifically, Fig. \ref{Flowchart} presents the proposed online identification and control structure.  
\color{black} The stability of such Koopman-inspired identification and control is guaranteed, which is proved in what follows. 

\subsection{ Stability Analysis} 

\color{black} MG dynamics can be expressed in a Koopman-based structure and can be approximated with the online Koopman-inspired identification in Section 3.3. The approximation error is bounded but often not quantifiable as it depends on the appropriateness of  Koopman observables and the online parameter identification algorithm. In what follows, we aim to prove stability properties in a general sense. \color{black}

\subsubsection{Proof of BIBO Stability} \label{Aderivation}
We prove that the proposed Koopman-inspired OKID-based control is BIBO (bounded-input-bounded-output) stable.
Denoted by $\hat{\bm{x}}_{k+1}$ the one-step-ahead prediction of the state vector $\bm{x}$ at the time step $k$ with the OKID-based estimation. Denoted by $\hat{\mathcal{K}}_k$ the estimated Koopman operator at time step $k$. According to Eq. (\ref{eq:edmd}), we have
\begin{small}
\begin{equation}
\begin{aligned}
\bm{g}(\bm{\hat{x}}_{k+1})= \bm{\Phi}(\bm{\hat{x}}_{k+1},\bm{0})= \hat{\mathcal{K}}_k \bm{\Phi}(\bm{x}_k,\bm{u}_k)
&=\hat{\mathcal{K}}_k \sum_{j=1}^{N_\varphi}\varphi_j({\bm{x}_k,\bm{u}_k})\bm{v}_j
 =\sum_{j=1}^{N_\varphi}(\rho_{j,k}\varphi_j(\bm{x}_k,\bm{u}_k)\bm{v}_j)
\end{aligned} \label{eq:edmd2}
\end{equation}\end{small}\noindent
where $\rho_{j,k}$ is the eigenvalue corresponding to the $j$th eigenfunction $\varphi_j$ for the estimated Koopman operator  $\hat{\mathcal{K}}_k$. 
Recall that $\bm{\Phi}(\bm{x},\bm{u})=\bm{g}(\bm{x})+\bm{l}(\bm{u})$ discussed in Section III.A, where $\bm{l}(\bm{u})=[{l}_1(\bm{u}),{l}_2(\bm{u}),…{l}_p(\bm{u})]^T$ and $\bm{l}(\bm{0})=\bm{0}$. \color{black} Then
\footnotesize
\begin{equation}
\begin{aligned}
\bm{g}(\bm{x}_{k+1})&=\bm{g}(\bm{\hat{x}}_{k+1}) + \bm{\delta}_{k} = \hat{\mathcal{K}}_k \bm{\Phi}(\bm{x}_k,\bm{u}_k) + \bm{\delta}_{k}\\
&=\hat{\mathcal{K}}_k (\bm{g}(\bm{x}_k)+\bm{l}(\bm{u}_k)) + \bm{\delta}_{k}\\
&=\hat{\mathcal{K}}_k (\hat{\mathcal{K}}_{k-1} \bm{\Phi}(\bm{x}_{k-1},\bm{u}_{k-1}) +\bm{\delta}_{k-1} +\bm{l}(\bm{u}_k)) + \bm{\delta}_{k}\\
&=\hat{\mathcal{K}}_k (\hat{\mathcal{K}}_{k-1}(\bm{g}(\bm{x}_{k-1}) + \bm{l}(\bm{u}_{k-1}))+  \bm{\delta}_{k-1}+\bm{l}(\bm{u}_{k})) + \bm{\delta}_{k}\\
&= \hat{\mathcal{K}}_k (\hat{\mathcal{K}}_{k-1}(\hat{\mathcal{K}}_{k-2} \bm{\Phi}(\bm{x}_{k-2},\bm{u}_{k-2}) + \bm{\delta}_{k-2} +\bm{l}(\bm{u}_{k-2}))  + \bm{l}(\bm{u}_{k-1})) +  \bm{\delta}_{k-1} + \bm{l}(\bm{u}_{k})) + \bm{\delta}_{k}\\
&=\dots= \prod_{h=0}^k \hat{\mathcal{K}}_{k-h}\bm{\Phi}(x_0,u_0)  + \sum_{h=1}^{k}\prod_{i=h}^k\hat{\mathcal{K}}_{k-i+h}(\delta_{h-1} + \bm{l}(\bm{u}_h))+\bm{\delta_{k}} \\
&=\sum_{j=1}^{N_\varphi}(\prod_{h=0}^{k}\rho_{j,h})\varphi_j(\bm{x}_0,\bm{u}_0)\bm{v}_j +\sum_{h=1}^{k} \prod_{i=h}^k\hat{\mathcal{K}}_{k-i+h} (\bm{\delta}_{h-1}+ \bm{l}(\bm{u}_h))+\bm{\delta}_{k}\nonumber
\label{ieq: Koopman1}
\end{aligned} 
\end{equation}\noindent\normalsize
\color{black}where $\bm{\delta} _{k}$ is the Koopman modeling error which has been defined in Eq. (\ref{eq:process}), and $\bm{v}_j$ is
the $j$th Koopman mode associated with the Koopman eigenfunction $\varphi_j$. Apparently,
\vspace{-0.1in}
\footnotesize
\begin{equation}
0 \leqslant
\|\sum_{j=0}^{N_\varphi}(\prod_{j=0}^{k}\rho_{j,h})\varphi_j(\bm{x}_0,\bm{u}_0)\bm{v}_j\|_2 \leqslant  \lim_{k\to\infty}(\mbox{max}_{j,h}|\rho_{j,h}|)^{k+1} \sum_{j=1}^{N_\varphi}\|\varphi_j(\bm{x}_0,\bm{u}_0)\bm{v}_j\|_2
\label{ieq:Koopmanineq}
\end{equation}\noindent
\normalsize

With LQR in the Koopman invariant subspace, assume the MG secondary controller can optimally make the magnitudes of all system eigenvalues smaller than 1 (if the system is stabilizable). That is $\hat{\mathcal{K}}_{k}{\varphi}_j = \rho_{j,k} \varphi_{j} $ with $|\rho_{j,k}|<1$. Due to the online \color{black}rolling-based \color{black} estimation in the proposed method, we can assume the global error $\|\sum_{h=1}^k\Pi_{i=h}^k\hat{\mathcal{K}}_{k}(\delta_{h-1}+l(u_h))\|_2$ is bounded by $\zeta_g$, and the modeling error is $\|\delta_{k}\|_2$ bounded by $\epsilon_m $. \color{black} According to (\ref{ieq:Koopmanineq}), we have 
\footnotesize
\begin{equation}
\hspace{-0.1in}\lim_{{k\to\infty}}\|\bm{g}(\bm{x}_{k+1})\|_2 
\leqslant  \lim_{{k\to\infty}}(\mbox{max}_{j,h}|\rho_{j,h}|)^{k+1} \sum_{j=1}^{N_\varphi}\|\varphi_j(\bm{x}_0,\bm{u}_0)\bm{v}_j\|_2  + \zeta_g  + \lim_{{k\to\infty}}\|\delta_{k}\|_2 
\leqslant  \zeta_g + \epsilon_m
\label{eq:errbounds}
\end{equation}\noindent
\normalsize
Based on (\ref{eq:errbounds}), $\bm{g}(\bm{x})$ converges till reaching the area  $\Xi = \big\{ \bm{g}(\bm{x}) | \|\bm{g}(\bm{x})\|_2 \leqslant \zeta_g + \epsilon_m \big\} $. 
Thus, the system is BIBO stable. Besides, the Koopman-based LQR can guarantee asymptotic stability subject to the disturbance in control input channels under mild conditions. See Section 3.5.2.

\subsubsection{Stability Margins of Koopman-Enabled LQR} \label{Aderivation1}
The discrete-time LQR used in this paper has analytical disc stability margins \cite{dLQRmargin}, within which asymptotic stability subject to the disturbance in control input channels is guaranteed. Specifically, consider the identified Koopman state space model described as below: 
\begin{small}
\begin{equation}
\begin{aligned}
\bm{g}(\bm{x}_{k+1})&=\bm{A}\bm{g}(\bm{x}_k) + \bm{B}\bm{u}_k+\bm{B}\bm{M}\bm{u}_k= \bm{A}\bm{g}(\bm{x}_k)+\bm{B}\bm{Kg}(\bm{x}_k)+\bm{B} \bm{M}\bm{Kg}(\bm{x}_k)\\
& = \bm{A}\bm{g}(\bm{x}_k)+\bm{B}(\bm{I}+\bm{M})\bm{Kg}(\bm{x}_k)
\end{aligned} \label{eq:ssmodelcomplex}
\end{equation}\end{small}\noindent
where $\bm{M}=diag([m_1, m_2, …m_{2N_{DER}}])$ is an introduced diagonal matrix to represent model uncertainty in control input channels. In other words, the introduced matrix parameter $\bm{M}$ can be used to quantify the uncertainty from control input channels, whereby one can provide the stability analysis based on the disc margin for each channel (which will be provided below). $\bm{K}$ is the control gain matrix such that $\bm{u}_k=\bm{Kg}(\bm{x}_k)$ in line with the LQR.

\begin{figure}
\centering
  \includegraphics[width=0.55\linewidth]{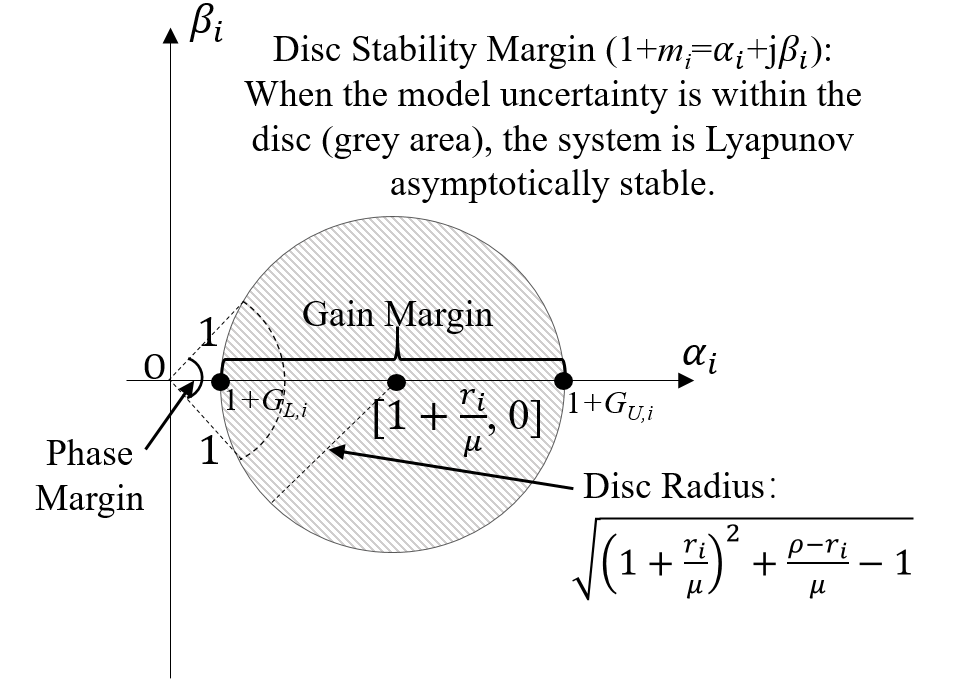}
    \caption{Disk stability margin for the discrete-time LQR}
  \label{fig: diskmargin}\end{figure}

Consider $\bm{M}$ and $\bm{g}(\bm{x}_k)$ to be complex-valued to reflect both gain and phase disturbances. \color{black} Define a Lyapunov function $ V(x)= \bm{g(x)}^{*}\bm{S}\bm{g(x)}$ (where $\bm{S}$ is the solution of Riccati equation). Based on the Lyapunov function and following the steps in \cite{dLQRmargin}, we provide the disk stability margin for the $i$th control input channel in (\ref{eq:disk}) without further explanation (also see Fig. \ref{fig: diskmargin}). Interested readers can refer to \cite{dLQRmargin} for the derivation of the disc margin.\color{black} 

\begin{small}
\begin{equation}
\begin{aligned}
1+m_i &= \bigg\{ \alpha_i + j\beta_i: \Big(\alpha_i-(1+\frac{r_i}{\mu})\Big)^2 + \beta_i^2 \\
&< (1+\frac{r_i}{\mu})^2 + \frac{\rho-r_i}{\mu}-1 \bigg\},  \quad \mbox{where} \quad i =1,2,...,2N_{DER}
\end{aligned} \label{eq:disk}
\end{equation}\end{small}\noindent
where $\rho=\sigma_{min}[\bm{Q}]/(\sigma_{max}[\bm{K}])^2$ and $\mu=\sigma_{max}[\bm{B}^T\bm{S}\bm{B}]$. $\sigma_{max}[.]$ and $\sigma_{min}[.]$ represent the matrix operation to obtain the maximum and minimum singular values, respectively. $r_i$ is the $i$th diagonal element of the cost matrix $\bm{R}$. 
Fig. \ref{fig: diskmargin} shows the disc margin, \color{black} within which the system is asymptotically stable. 
Specifically, 
according to Eq. (\ref{eq:disk}) and 
Fig. \ref{fig: diskmargin}
, the sufficient conditions of asymptotic convergence against the model uncertainty is\color{black}: $1+G_{L,i} <\alpha_i<1+G_{U,i}$ for the gain margin and $PM_{L,i} <\arctan\frac{\beta_i}{\alpha_i}<PM_{U,i}$ for the phase margin, with 
\begin{small}
\begin{equation}
\begin{aligned}
G_{L,i} = \frac{r_i}{\mu}-\sqrt{(1+\frac{r_i}{\mu})^2+\frac{\rho-r_i}{\mu}-1},\quad
G_{U,i} = \frac{r_i}{\mu}+\sqrt{(1+\frac{r_i}{\mu})^2+\frac{\rho-r_i}{\mu}-1} \\
\end{aligned} \label{eq:mbounds}
\end{equation}\end{small}\noindent
and
\begin{small}
\begin{equation}
\begin{aligned}
PM_{L,i} &= -\arccos(1/(1+\frac{r_i}{\mu})) = -\arccos \frac{\mu}{\mu+r_i}\\
PM_{U,i}&= \arccos(1/(1+\frac{r_i}{\mu})) =\arccos \frac{\mu}{\mu+r_i}, &\mbox{for} \quad i = 1,2,...,2N_{DER}.
\end{aligned} \label{eq:mbounds}
\end{equation}\end{small}\noindent

\color{black}

\section{Case Studies}\label{sec:case studies}
\begin{figure}[!b]
\centering
  \includegraphics[width=0.78\linewidth]{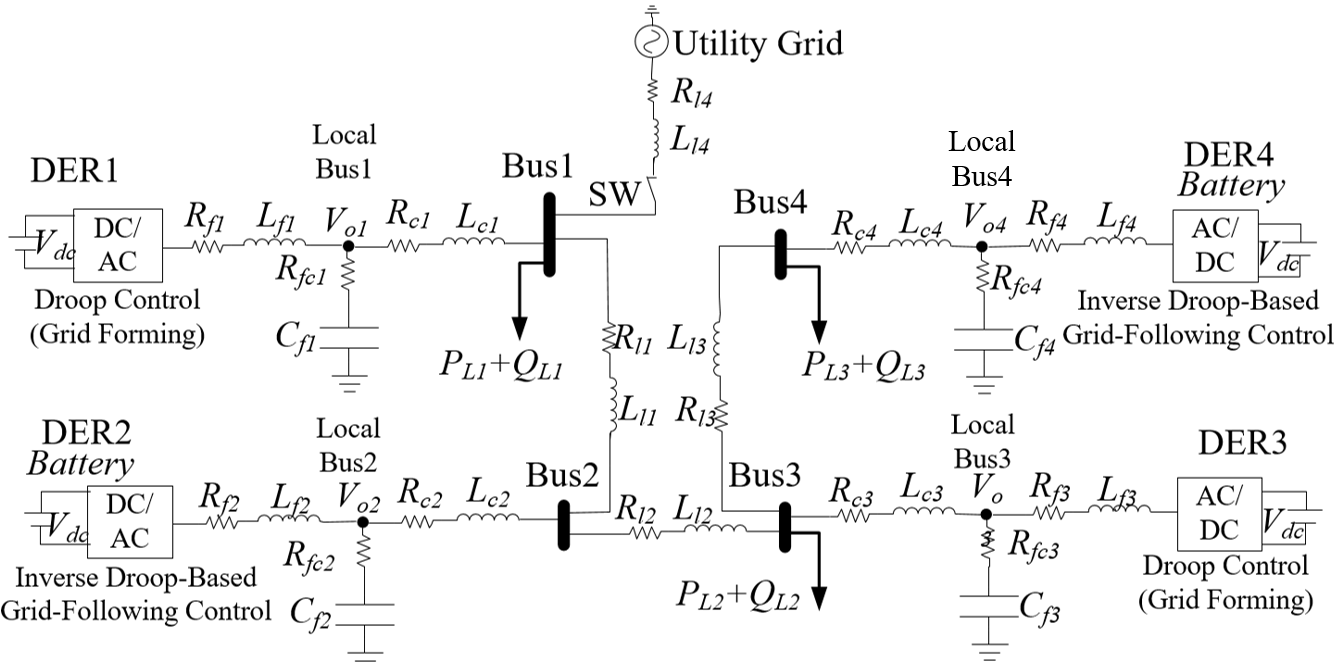}\\
\caption{The MG 4-bus test system}
  \label{fig: MG4bus}\end{figure}

\begin{figure}[!b]
\centering
  \includegraphics[width=0.8\linewidth]{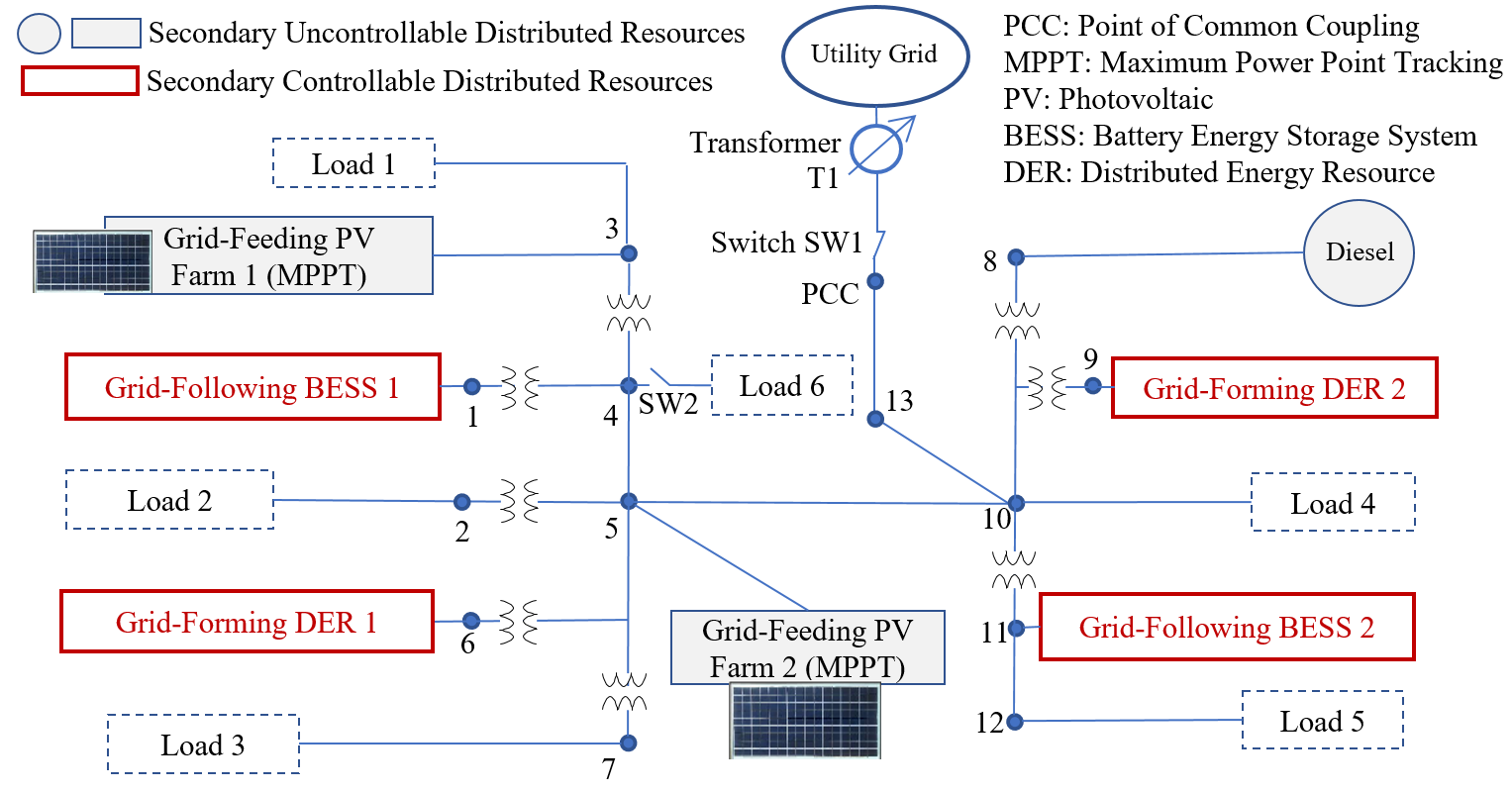}\\
\caption{The MG 13-bus test system}
  \label{fig: MG13bus}\end{figure}

This section presents case studies based on two MG test systems, namely a four-bus MG as shown in Fig. \ref{fig: MG4bus} and a thirteen-bus MG as shown in Fig. \ref{fig: MG13bus}, to verify the effectiveness of the proposed Koopman-inspired identification and control. The two test systems were established in MATLAB Simulink 2021b. The DERs in the test systems are primary-controlled in different control modes (grid-forming converters, grid-following converters, and an isochronous-controlled diesel generator as given in Fig. \ref{fig: MG4bus} and Fig. \ref{fig: MG13bus}) with the inner control loops modeled in detail. \color{black} Therefore, the interaction of primary and secondary control is preserved \color{black} in simulation to test the effectiveness of secondary control in realistic setups. The implementation of the converter voltage and current control inner-loops can be found 
\cite{Ma2021,6200347}. 
 
 \begin{table}[!ht]
\centering
\caption{Parameters of the 4-Bus MG Test System}
\label{tab:MG1}
\begin{tabular}{c c}
\hline
Parameters&Value\\
\hline
Power base $S_{base}$ & 30kVA\\
Voltage Base $V_{base}$ & 480V\\
Primary control time step $T_{sp}$ & 0.1ms\\
Secondary control time step $T_{s}$ & 30ms\\
Sliding  window length for estimation $N$ &  9 (270ms) \\
Local Voltage proportional gain $K_{P}$ & 0.5 \\
Local Voltage integral gain $K_{S}$ & 523 \\
Local current proportional gain $K_{P}$ & 0.3 \\
Local current integral gain $K_{S}$ & 635 \\
Frequency droop parameters for DERs 1,2:  $\sigma_{\omega}$ & $2.14\times 10^{-3} rad/(W \cdot s)$ \\
Voltage droop parameters for DERs 1:  $\sigma_{V}$ & $1.0\times 10^{-3} V/Var$  \\
Voltage droop parameters for DERs 2:  $\sigma_{V}$ & $6.3\times 10^{-3} V/Var$  \\
Frequency droop parameters for DERs 3,4:  $\sigma_{\omega}$ & $2.83\times 10^{-3} rad/(W \cdot s)$\\
Voltage droop parameters for DERs 3:  $\sigma_{V}$ & $1.5\times 10^{-3} V/Var$  \\
Voltage droop parameters for DERs 4:  $\sigma_{V}$ & $9.4\times 10^{-3} V/Var$  \\
PMU measurement noise & $\mathcal{N}(0,0.0056^2)$\\
Control Time delay & $\mathcal{N}(0.05,0.002^2) s$ \\
Ambient perturbation level added to the reference \\of DER output voltage and angle: & $\mathcal{N}(0,0.01^2)$\\
Filter resistance $R_{f1,2,3,4}(\Omega)$ & 0.1\\
Filter inductance $L_{f1,2,3,4},L_{c1,2}(mH)$ & 1.35\\
Filter capacitance $C_{f1,2,3,4}(\mu F)$ & 50\\
Filter capacitor resistance $R_{fc1,2,3,4}(\Omega)$ & 1\\
Line resistance $R_{c1,2}(\Omega)$ & 0.08\\
Line resistance $R_{c3,4}(\Omega)$ & 0.09\\
Line inductance $L_{c1,2}(mH)$ & 0.35\\
Line inductance $L_{c3,4}(mH)$ & 0.45\\
Line Resistance $R_{l1,2,3,4}(\Omega)$ & 0.15, 0.35, 0.23, 0.17\\
Line inductance $L_{l1,2,3,4}(mH)$ & 0.42, 0.33, 0.55, 2.40\\
Load $P_{L1,2,3}$ (active power in kW) & 20, 16, 12\\
Load $Q_{L1,2,3}$ (reactive powe in kVar) & 9, 9, 6\\
LQR control parameter $q_{V}$ & $1\times 10^3\bm{I}$\\
LQR control parameter $q_{\sin}$ , $q_{\cos}$ & $\bm{0}$\\
LQR control parameter $q_{\omega}$ & $1\times 10^{-6}\bm{I}$\\
LQR control parameter ${r_P}$, ${r_Q}$ & $1\times 10^{-6}\bm{I}$\\
Control input lower bounds $U_{LB}$  & -1.0 kVA\\
Control input upper bounds $U_{UB}$ & 1.0 kVA\\
Time period for the optimization (\ref{eq:OKID-ERA-Implement2}) $\quad T_{OPT}$ & 0.6 s\\
\color{black} Time constant of the power low-pass filter & \color{black}0.02857s\\

\hline
\end{tabular}
\begin{tablenotes}
\small \item * $\mathcal{N}(a,b)$ is the normal distribution with mean of $a$ and variance of $b$. 
Control parameters are designed based on Per Unit.
\end{tablenotes}\end{table}

Besides, randomized measurement noises, control time delays, and ambient perturbations were incorporated into the test systems to mimic practical operation. The simulation parameters of the two test systems are summarized in Table \ref{tab:MG1} and Table \ref{tab:MG2}, respectively. \color{black} The readers can find more information about the test systems at https://github.com/nash13123/MG-Test-System.git\color{black}.

\begin{table}[!t]
\centering
\caption{Parameters of the 13-Bus MG Test System}
\label{tab:MG2}
\begin{tabular}{c c}
\hline
Parameters&Value\\
\hline
Power base $S_{base}$ & 150kVA\\
Voltage Base $V_{base}$ & 4.16kV\\
Sliding  window length for estimation $N$ &  14(420ms) \\
Droop parameters for all DERs:  $\sigma_{\omega}$ & $3.14\times 10^{-4} rad/(W \cdot s)$ \\
Droop parameters for all DERs:  $\sigma_{V}$ & $1.5\times 10^{-3} V/Var$  \\
Ambient perturbation level 
& $\mathcal{N}(0,0.02^2)$\\
LQR control parameter $q_{V}$ & $1\times 10^3\bm{I}$\\
LQR control parameter $q_{sin}$ , $q_{cos}$ & $\bm{0}$\\
LQR control parameter $q_{\omega}$ & $0.01\bm{I}$\\
LQR control parameter ${r_P}$, ${r_Q}$ & $1\times 10^{-6}\bm{I}$\\ 
\hline
\end{tabular}
\begin{tablenotes}
\small \item * 
Other control parameters are the same to the values in Table \ref{tab:MG1}.
\end{tablenotes}
\end{table}

\subsection{Identification and Control in the 4-Bus MG Test System}

The small 4-bus MG test system was used to test the proposed Koopman-inspired enhanced OKID with control under load variations and the MG transition from the grid-connected mode to islanded mode. The DERs at  Bus 1 and 3 are droop-based grid-forming, and the DERs at Bus 2 and 4 are inverse-droop-based grid-following. At 0.7s, the MG was disconnected from the main grid by turning off the switch SW, \color{black} which causes \color{black} sudden voltage drops and consequent dynamics. After detecting the sudden change, the secondary control was enabled and kept online from 0.8s, i.e., approximately 0.1s lag to mimic a time delay of islanding event detection in practical applications. 

\begin{figure}[!t]
\centering
\includegraphics[width=0.75\linewidth]{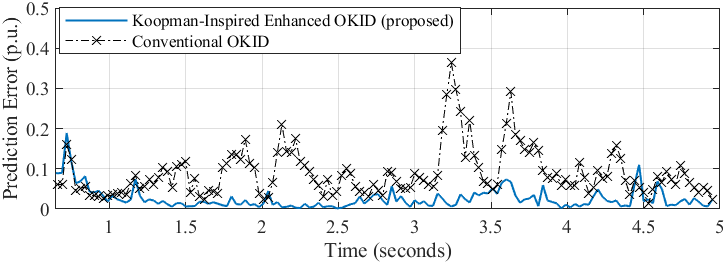}
\caption{Comparison of the prediction error}
  \label{fig: prederr}\end{figure}

\textbf{\emph{Modeling accuracy of the Koopman-inspired OKID.}}
First, we evaluate the modeling accuracy with the one-step-ahead prediction error of the voltage magnitude, which is defined as
\begin{small}
\begin{equation}
\begin{aligned}
\bm{e}_{k+1}^{(pred)} 
= \frac{1}{dim{(\bm{\Delta V})}}\|\bm{\Delta V_{k+1}}-\bm{\Delta \hat{V}_{k+1}}\|	
 \end{aligned} \label{eq:LQRcost}
\end{equation}
\end{small}\noindent
where $dim[.]$ represents the dimension of the vector in the bracket, and $\Delta\hat{\bm{V}}_{k+1}$ represents the predicted voltage magnitude at time step $k+1$ by the identified model 
of interest. 
\color{black} 
In Fig. \ref{fig: prederr}, we compared the prediction error of two different ways of modeling: 
(i) the  proposed Koopman-inspired  enhanced OKID with the basis $\bm{z}=[\Delta \bm{V}, \sin\bm{\theta}-\sin(\bm{\theta}_L^*), \cos\bm{\theta}-\cos(\bm{\theta}_L^*), \Delta\bm{\omega}]^T$; 
(ii) the conventional  OKID (i.e., linearize the system model in Eq. (\ref{nldroopgf}) and apply OKID with $\gamma_{opt}$ fixed at $\frac{1}{2}$). 
It was found that the proposed Koopman-inspired enhanced OKID leads to smaller prediction error than the conventional OKID. These results show that the salient features of the proposed Koopman-inspired OKID, i.e., the Koopman nonlinear basis  
and the adaptive $\gamma_{opt}$, 
can ensure a good modeling accuracy regardless of nonlinearity and uncertainty during large disturbances. 

\begin{figure}[!ht]
\centering
  \includegraphics[width=0.42\linewidth]{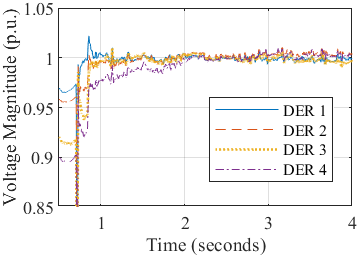} \qquad
  \includegraphics[width=0.42\linewidth]{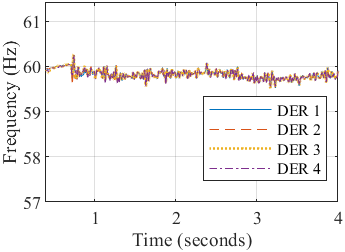}\\
 (a) The proposed Koopman-inspired enhanced OKID with LQR control\\
    \includegraphics[width=0.42\linewidth]{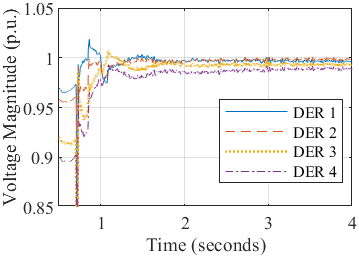}\qquad
  \includegraphics[width=0.42\linewidth]{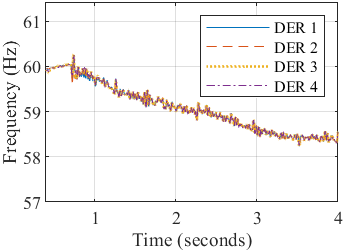}\\
  (b) The secondary PI control\\
    \includegraphics[width=0.42\linewidth]{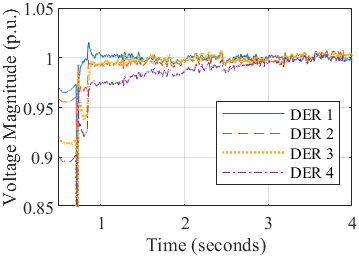}\qquad
  \includegraphics[width=0.42\linewidth]{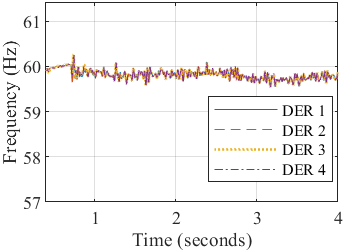}\\
(c) The conventional OKID with LQR control \color{black} ($\gamma = 0.5$)\\ 
 \includegraphics[width=0.43\linewidth]{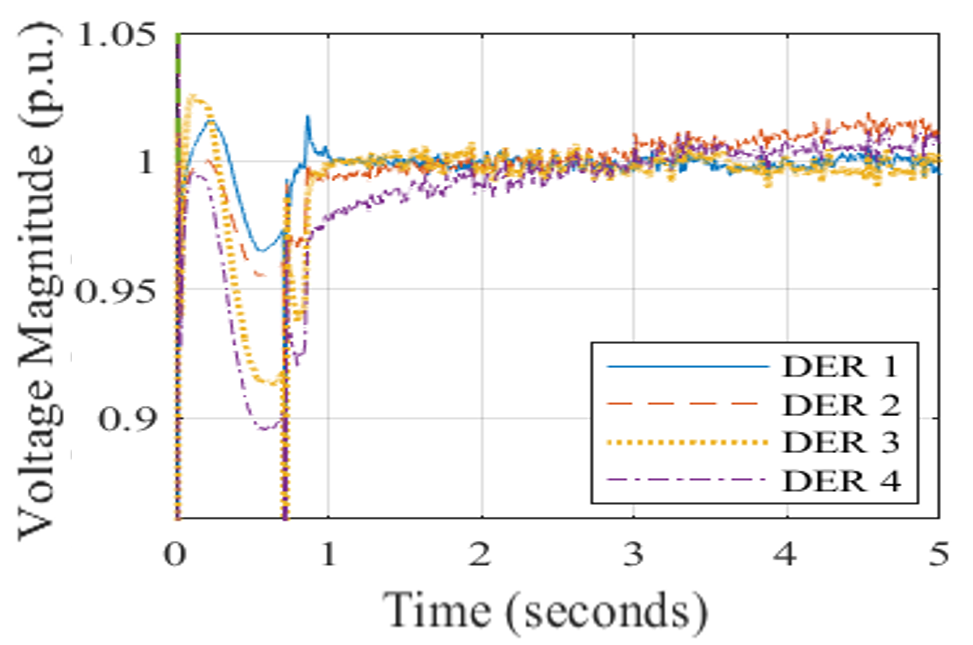} \quad
  \includegraphics[width=0.43\linewidth]{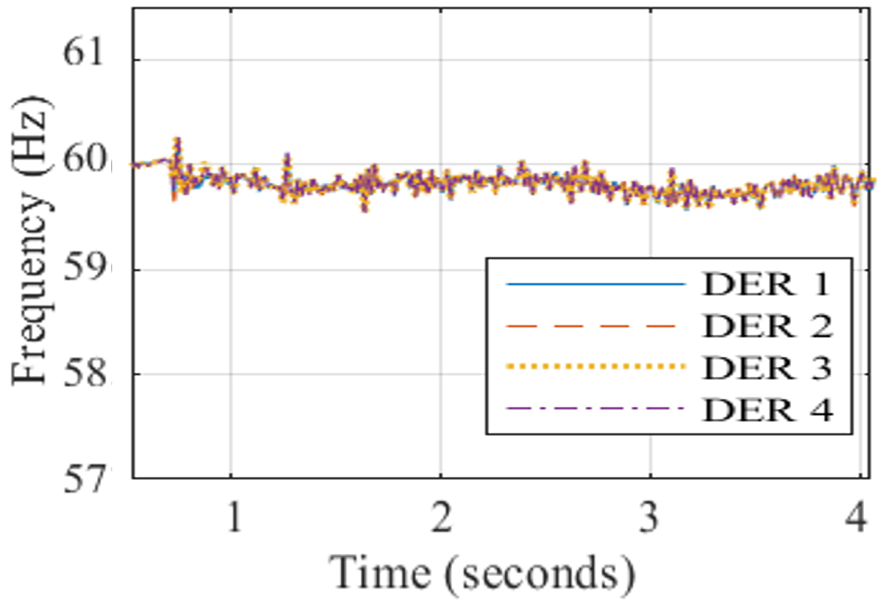}\\
  (d) The classical EDMDc (least-squares-based Koopman operator control with the Koopman observables $\bm{z}$) 
  \caption{\color{black}Voltage and frequency trajectories of the 4-bus MG test system with different secondary control methods}
  \label{fig: MG4Results}\end{figure}

\textbf{\emph{Control results comparison.}} Fig. \ref{fig: MG4Results} compares the voltage and frequency trajectories with different secondary control methods. 
As Fig. \ref{fig: MG4Results}(a)-(c) show, the bus voltage suddenly drops with incurred transients after the disturbance at 0.7s, which triggers the secondary control to restore the voltage and frequency to their nominal values (1p.u and 60Hz).
Fig. \ref{fig: MG4Results}(a) shows the control results with the proposed Koopman-inspired enhanced OKID with LQR control; 
both voltage and frequency are corrected approximately to the nominal values. 
For comparison, Fig. \ref{fig: MG4Results}(b) presents the voltage and frequency trajectories using secondary PI control that is tuned with the best effort.
The PI control with respect to the voltage magnitude and the frequency shows a slower response for voltage restoration compared to the proposed control, and has non-zero steady-state errors. It also suffers from a larger frequency deviation as it cannot handle 
the voltage-frequency dependence properly. 
Fig. \ref{fig: MG4Results}(c) shows the results of conventional OKID with LQR control. In contrast to the proposed method, 
the conventional OKID with LQR cannot realize the same fast voltage restoration.
\begin{figure}[b]
\centering
 \includegraphics[width=0.45\linewidth]{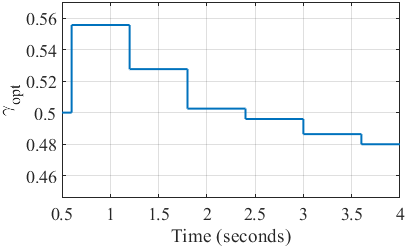}\quad
\caption{\color{black} Estimated $\gamma_{opt}$ of the proposed Koopman-inspired enhanced OKID}
\label{fig: gammapot}\end{figure}
\begin{figure}[!bt]
\centering
\includegraphics[width=0.45\linewidth]{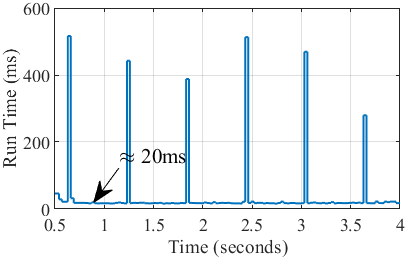}\\
\caption{\color{black} Run time of the proposed Koopman-inspired enhanced OKID}
  \label{fig: runtime}\end{figure}
Fig. \ref{fig: MG4Results}(d) shows the voltage and frequency trajectories of the classical EDMDc (i.e., LQR with pseudo-inverse least-squares identification based on the Koopman observables $\bm{z}$). By comparing Fig. \ref{fig: MG4Results}(d) with Fig. \ref{fig: MG4Results}(a), we found that the LQR with the least-squares-based identification cannot perform as well as the LQR with the proposed Koopman-inspired Koopman-inspired identification. \color{black}
These results indicate the effectiveness of the proposed Koopman-inspired enhanced OKID that possibly results from the two ingredients: (i) the nonlinear basis functions of the Koopman observables proposed in Eq. (\ref{eq:bases}); (ii) the OKID with adaptive $\gamma_{opt}$. Both ingredients help better describe the MG systemwide dynamics under big disturbances, realizing more effective control for both voltage and frequency. 

Fig. \ref{fig: gammapot} presents the optimized parameters $\gamma_{opt}$ during control, and Fig. \ref{fig: runtime} shows \color{black}the \color{black} run time of the proposed Koopman-inspired enhanced OKID at each time step of secondary control. The run time of the proposed method is about 20ms \color{black} in case that $\gamma_{opt}$ is not updated\color{black}, less than the time step of secondary control (30ms). The run time of the proposed method is around 250-500ms 
\color{black} in case that $\gamma_{opt}$ is updated\color{black}, which is still less than the time period $T_{OPT}=0.6s$ between two updates of $\gamma_{opt}$. These indicate the feasibility to implement the proposed Koopman-inspired identification and control online. 
\begin{figure}[!ht]
\centering
  \includegraphics[width=0.43\linewidth]{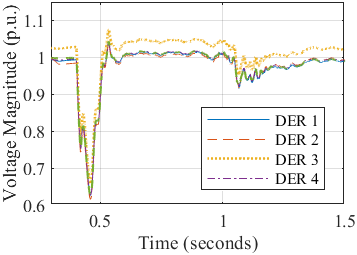} \qquad
  \includegraphics[width=0.43\linewidth]{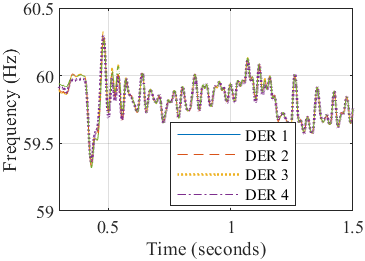}  \\
  (a) The proposed Koopman-inspired enhanced OKID with LQR control\\ 
   \includegraphics[width=0.43\linewidth]{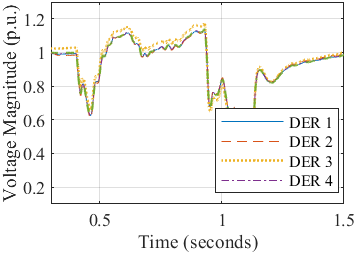} \qquad 
   \includegraphics[width=0.43\linewidth]{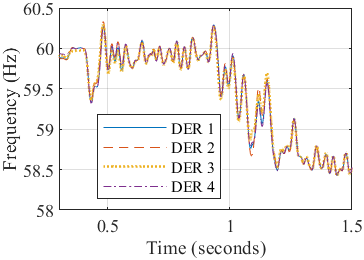}\\
  (b) The secondary PI control\\
    \includegraphics[width=0.43\linewidth]{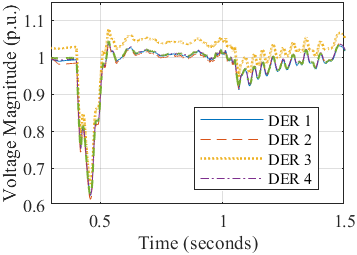} \qquad
    \includegraphics[width=0.43\linewidth]{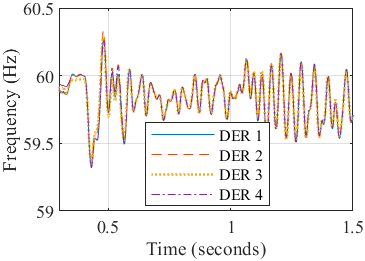}\\
 (c) The conventional OKID with LQR control \color{black}($\gamma = 0.5$)\\ 
\includegraphics[width=0.47\linewidth]{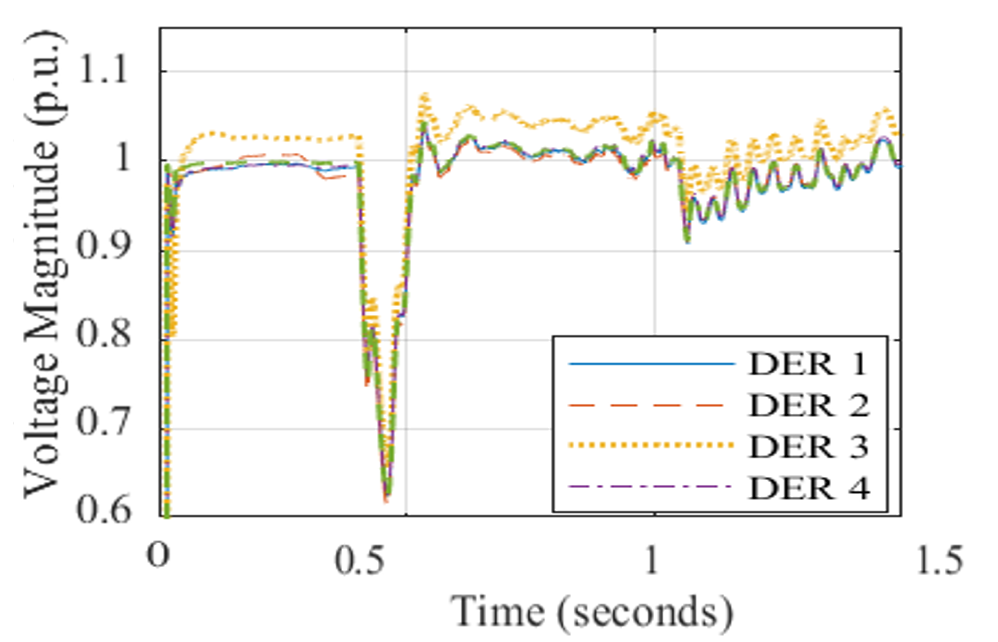}
\includegraphics[width=0.48\linewidth]{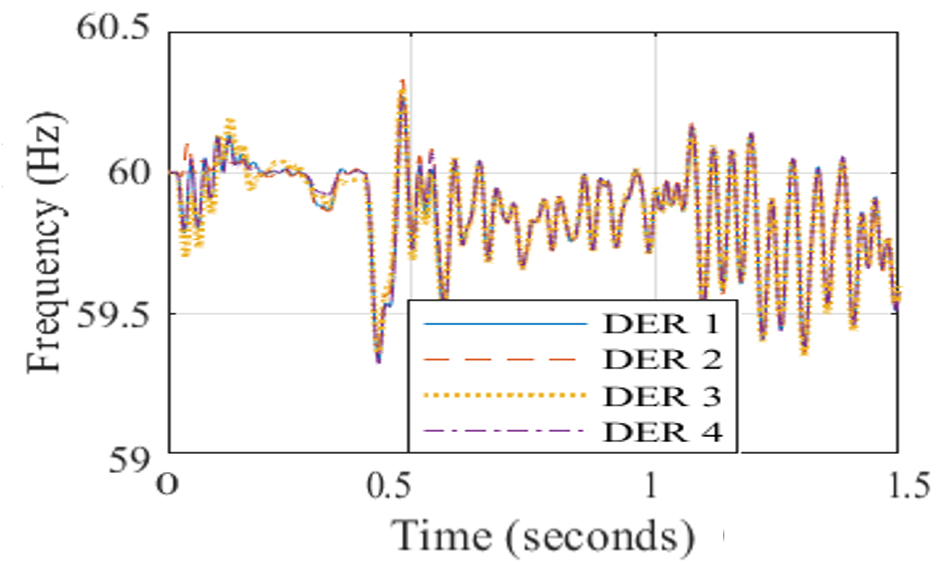}\\
 (d) The classical EDMDc (least-squares-based Koopman operator control with the Koopman observables $\bm{z}$)
  \caption{Voltage and frequency trajectories of the 13-bus MG test system with different secondary control methods}\label{fig: MG13busResults}\end{figure}

\color{black} 

\color{black}

\subsection{Identification and Control in the 13-Bus MG Test System}
To show the performance of the Koopman-inspired enhanced OKID with LQR control in larger systems for generality, 
we consider the 13-bus MG test system presented in Fig. \ref{fig: MG13bus}, \color{black} which is adapted from the IEEE 13-node test feeder \cite{testfeeders}
. \color{black} The DERs at Bus 6 and 9 are droop-based grid-forming. The 
BESSs at Bus 1 and 11 are inverse-droop-based grid-following. The solar farms at Bus 3 and 5 are grid-following under MPPT, which are not controllable for secondary control. \color{black}The MG system is under transition from the grid-connected to the islanded modes, and under generation/load variations. \color{black} At 0.4s, the MG is disconnected from the main grid by turning off the switch SW1, causing the sudden drop of voltage with incurred transient. 
After detecting the islanding transient, the secondary control is triggered and kept online from 0.5s, i.e., 0.1s lag to mimic a time delay of islanding detection in practical application. Next, an active power perturbation of the two solar farms (around 80kW for each) occurs at 1.0s due to a drop of the solar irradiation from 1000 to 200 $W/m^2$. 
\color{black} Then, a load perturbation happens at the Bus 4 at 1.05s: the consumed active power increases by 150kW and the consumed reactive power increased by 50kVar by turning on the switch SW2. 

Fig. \ref{fig: MG13busResults} compares the voltage and frequency trajectories with different secondary control methods. 
 Fig. \ref{fig: MG13busResults}(a) shows that the proposed Koopman-inspired OKID with LQR control can correct the voltage and frequency approximately to the nominal values.
 For comparison, the voltage and frequency trajectories with the secondary PI control are shown in Fig. \ref{fig: MG13busResults}(b),
which illustrates that the PI control with the best effort of tuning still fails to realize the stable and accurate voltage and frequency restoration.
Fig. \ref{fig: MG13busResults}(c) shows the results of 
the conventional OKID with LQR control, which  
suffers from larger voltage and frequency oscillations after 1.0s.
\color{black} Fig. \ref{fig: MG13busResults}(d) shows the voltage and frequency trajectories of classical EDMDc (i.e., LQR with pseudo-inverse least-squares identification based on the Koopman observables $\bm{z}$). By comparing Fig. \ref{fig: MG13busResults}(d) with Fig. \ref{fig: MG13busResults}(a), we found that the classical EDMDc cannot perform as well as the LQR with the proposed Koopman-inspired Koopman-inspired identification. \color{black}
These results further demonstrate the advantages of the proposed Koopman-inspired OKID with LQR control. Because of the nonlinear Koopman embeddings and the adaptive $\gamma_{opt}$, the proposed method can effectively restore the voltage and frequency to their nominal values despite nonlinearity and uncertainty due to large disturbances.  


\section{Conclusions}
This paper proposed a data-driven Koopman-inspired identification and control method for MG secondary voltage and frequency control. \color{black} The proposed method requires \color{black} no knowledge of network information and primary controllers. It requires no warm-up training yet with guaranteed BIBO stability and even \color{black} asymptotic \color{black} stability under some mild conditions. In this method, a Koopman operator-inspired enhanced OKID (observer Kalman filter identification) algorithm is proposed, whereby the Koopman state space model is estimated online and used for control to handle microgrid nonlinearity and uncertainty adaptively. Case studies in the 4-bus and 13-bus MG test systems (with different converter control modes) demonstrate the effectiveness and robustness of the proposed Koopman-inspired identification and control method subject to mode transitions, varying operating conditions, measurement noises and time delays.
\bibliographystyle{elsarticle-num}
\bibliography{mybib.bib}












\end{document}